%% file: ms.tex
\documentclass{article}

\usepackage{microtype}
\usepackage{graphicx}
\usepackage{subfigure}
\usepackage{booktabs} 

\usepackage{hyperref}

\usepackage[accepted]{icml2023}

\usepackage{amsmath}
\usepackage{amssymb}
\usepackage{mathtools}
\usepackage{amsthm}

\usepackage[capitalize,noabbrev]{cleveref}

\usepackage{mathtools}
\usepackage{enumerate}
\usepackage{amssymb, amsfonts}
\usepackage{relsize}
\usepackage[american]{babel}
\usepackage{multirow}
\usepackage{aligned-overset}

\theoremstyle{plain}
\newtheorem{theorem}{Theorem}[section]
\newtheorem{proposition}[theorem]{Proposition}

\theoremstyle{definition}

\newtheorem{assumption}[theorem]{Assumption}
\theoremstyle{remark}
\newtheorem{remark}[theorem]{Remark}
\newtheorem{example}[theorem]{Example}

\newcommand{\argdot}{\,\cdot\,}

\DeclareMathOperator*{\argmax}{arg\,max}
\DeclarePairedDelimiter\abs{\lvert}{\rvert}%
\newcommand{\norm}[1]{\left\lVert#1\right\rVert}
\allowdisplaybreaks

\usepackage[textsize=tiny]{todonotes}

\usepackage{xcolor}
\definecolor{blue}{RGB}{28,132,218}
\definecolor{orange}{RGB}{237,125,49}
\definecolor{darkviolet}{rgb}{0.58, 0.0, 0.83}

\icmltitlerunning{Enabling First-Order Gradient-Based Learning for Equilibrium Computation in Markets}

\begin{document}

\twocolumn[
\icmltitle{Enabling First-Order Gradient-Based Learning\\for Equilibrium Computation in Markets}


\icmlsetsymbol{equal}{*}

\begin{icmlauthorlist}
\icmlauthor{Nils Kohring}{sch}
\icmlauthor{Fabian R.\ Pieroth}{sch}
\icmlauthor{Martin Bichler}{sch}
\end{icmlauthorlist}

\icmlaffiliation{sch}{School of Computation, Information and Technology, Technical University of Munich}

\icmlcorrespondingauthor{Nils Kohring}{\href{mailto:nils.kohring@tum.de}{nils.kohring@tum.de}}


\vskip 0.3in
]



\printAffiliationsAndNotice{}  

\begin{abstract}
    Understanding and analyzing markets is crucial, yet analytical equilibrium solutions remain largely infeasible. Recent breakthroughs in equilibrium computation rely on zeroth-order policy gradient estimation. These approaches commonly suffer from high variance and are computationally expensive.
	The use of fully differentiable simulators would enable more efficient gradient estimation. However, the discrete allocation of goods in economic simulations is a non-differentiable operation. This renders the first-order Monte Carlo gradient estimator inapplicable and the learning feedback systematically misleading. 
	We propose a novel smoothing technique that creates a surrogate market game, in which first-order methods can be applied.
	We provide theoretical bounds on the resulting bias which justifies solving the smoothed game instead. These bounds also allow choosing the smoothing strength a priori such that the resulting estimate has low variance. Furthermore, we validate our approach via numerous empirical experiments.
	Our method theoretically and empirically outperforms zeroth-order methods in approximation quality and computational efficiency.
\end{abstract}

\section{Introduction}

Auctions are at the center of modern economic theory. Given some private valuation of goods available for purchase, participants must place bids on the market that maximize their expected payoff while remaining unaware of the other participants' valuations.
In the seminal paper \cite{vickrey1961counterspeculation} the foundation for most auction theory results of today was laid. 
It is crucial to understand the strategic behavior in various auction applications, ranging from treasury and industrial procurement auctions to spectrum sales.
Depending on the circumstances and behavioral assumptions, optimal strategies may differ drastically, starting from strategies, such as understating demand (bid-shading) \cite{krishna2009auction} and overstating demand (overbidding) \cite{ott2013incentives}, or much more convoluted strategies. However, computing such equilibria and approximations a priori remains challenging. Analytical equilibria can only be derived under strong assumptions such as in single-item auctions or the independent private values model.

A recent approach based on policy optimization uses randomized finite difference approximations of the gradient \cite{bichler2021learning}. They proposed an algorithm called \emph{neural pseudogradient ascent} (NPGA), which parametrizes the bidding strategies using neural networks and follows the approximate gradient dynamics of the game via simultaneous gradient ascent of all agents. The gradients are computed via \emph{evolution strategies} (ES) \cite{salimans2017evolution}, which smoothen the objective by adding noise in the parameter space, thereby treating the environment as a black box.
Compared with the well-known REINFORCE algorithm, where the actions are perturbed, this also results in zeroth-order gradient estimates with better precision and lower variance but much higher computational cost.

Under the differentiable programming paradigm, there is a growing interest in computing gradients for numerous reinforcement learning applications that allow for first-order gradient estimates. It is possible to create a full computational graph for applications with a certain amount of structure. 
First-order methods have the advantage of much lower variance, which leads to faster convergence rates to local minima of non-convex objective functions \cite{mohamed2020monte}.
However, there are two common problems in employing first-order methods.
First, most reinforcement learning environments are provided only as black boxes. This implies that there is no explicit access to the underlying state transition function and the gradient can only be estimated by repeatedly evaluating the reward function. 
The wide applicability of zeroth-order policy optimization, like REINFORCE and more advanced actor-critic techniques \cite{schulman2017proximal}, contributes to their popularity.
Second, in some applications, such as the training of variational autoencoders, the computational path of the derivate is blocked (i.e., repeatably applying the chain rule to calculate the gradient of the reward with respect to the parameters of the policy) because it consists of sampling a random variable, which is a non-differentiable operation \cite{bangaru2021systematically}.

The situation is similar in auction games. The allocation of indivisible goods causes biased gradients of first-order methods. Example~\ref{ex:naive-failure} showcases this observation. It was observed that the first-order Monte Carlo gradient estimate does not converge to equilibrium and quickly causes consistent zero-bidding \cite{bichler2021learning}.
From a mathematical standpoint, the single-sample (ex post) utility has a discontinuity. Thus, the sample mean of its exact gradients is an inadequate estimate for its true (ex ante) utility gradient (the expected utility over all possible valuations).

\begin{example}\label{ex:naive-failure}
	Consider a first-price sealed-bid (FPSB) single-item auction. Two bidders compete for a single good, where the winner pays his or her bid. The derivative of the utility with respect to the bid is zero for losing bids and minus one for winning bids after a point of discontinuity. Either the bidder loses and receives no feedback or wins and could have won with an even smaller bid.
\end{example}

\begin{figure}[t]
	\centering
	\includegraphics[width=0.85\columnwidth]{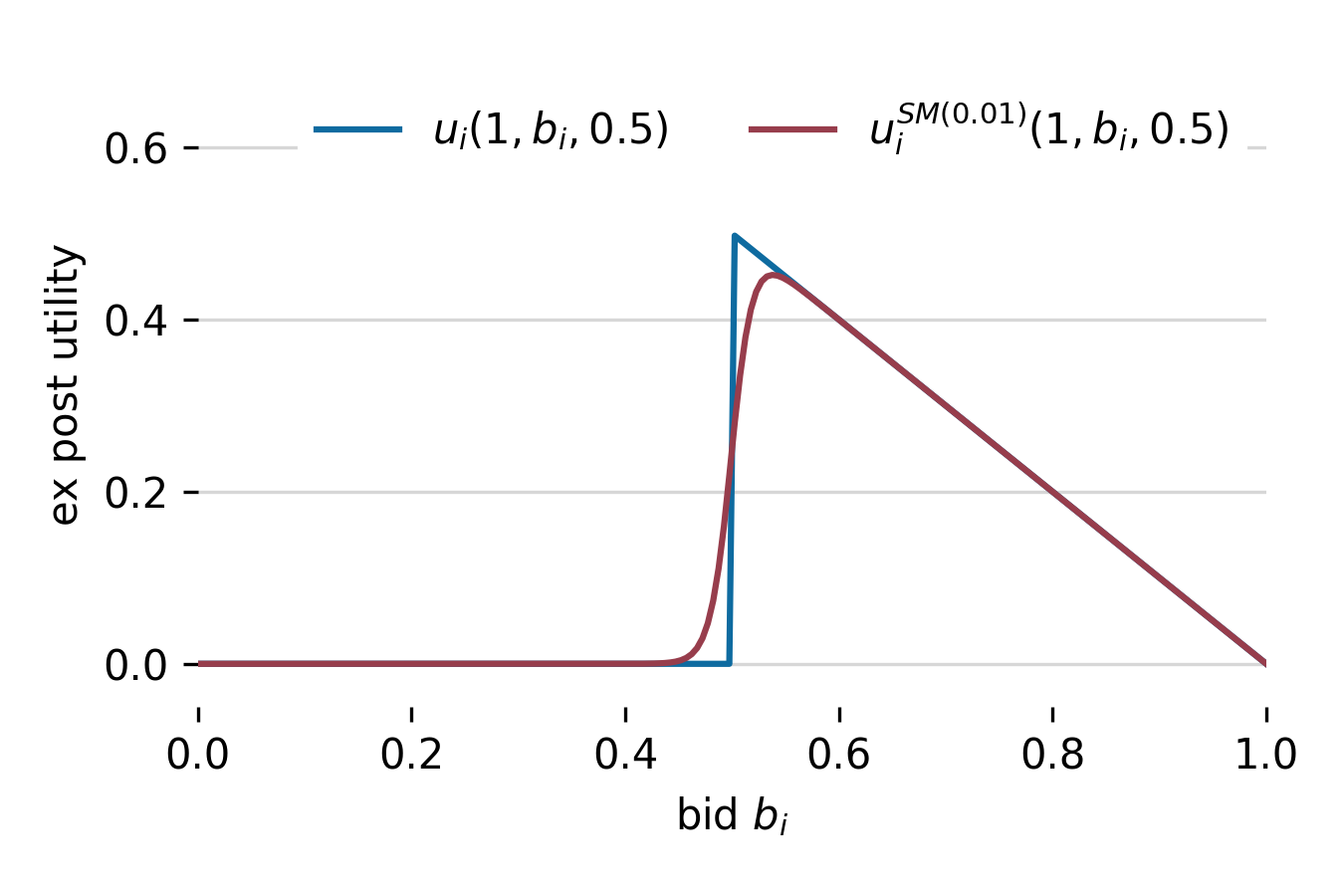}
	\caption{Ex post utility in the original FPSB single-item auction (blue) and its smoothed version (red) for a temperature of $\lambda = 0.01$ and a highest opponent bid of $0.5$. The utility in the original auction is zero for losing bids and decreases linearly for winning bids.}
	\label{fig:ex-post-utility}
\end{figure}

In this study, we propose transforming multi-agent auction games such that their utility functions are sufficiently regular for applying efficient first-order gradient methods while keeping the overall gradient dynamics close to the original game. In contrast to the original allocations of indivisible items, we use \emph{soft} allocations instead. We effectively treat the items as divisible and allocate the proportional fraction of an item to the bidders based on their reported bids. An additional adaption to the pricing rule eliminates the discontinuity at the threshold of winning and losing an object. However, this comes at the expense of introducing a bias in the utility function.
For example, a losing bidder has zero utility the original auction. However, in the smoothed auction, this bidder receives a small fraction of the good (and pays a correspondingly small price), such that the gradient indicates that a higher bid would have resulted in higher utility. The feedback to bid lower when winning remains of similar magnitude. Thus, there is always appropriate feedback on the current bidding strategy in the smoothed game. Figure~\ref{fig:ex-post-utility} shows the utility function and its relaxed version. 

This approach is applicable widely to economic models and general auction formats, such as sequential or simultaneous sales of multiple goods, as in combinatorial auctions with item bidding. It is further independent of the number of bidders, payment rule, risk preferences of the bidders, or correlations among the bidders' valuations. We demonstrate that the choice of a smoothing parameter follows a natural trade-off.
Importantly, computing equilibria in multi-agent games is not straightforward and many negative results are known \cite{chasnov2020convergence,mazumdar2020policy,letcher2020impossibility}. Therefore, changes to the game dynamics must be implemented with great caution, and we can prove that an approximate equilibrium in the smoothed game still constitutes an approximate equilibrium in the original game.

Computational-wise, the smoothing only comes with the cost of tracking the gradients of the individual operations, upon which the game dynamics are built. 
Compared with NPGA, learning in the \emph{smooth market} (SM) via the first-order estimator is more than ten times faster while yielding better results. For example, an iteration of NPGA in a small single-item auction with the default hyperparameters from \cite{bichler2021learning} takes approximately $0.16 \,$s, whereas first-order policy gradients applied to the SM take an average of $0.01 \,$s.

Our contribution can be summarized as follows: We introduce the SM and show that first-order methods provide an unbiased estimator of the utility gradient of the SM game. Furthermore, we provide theoretical guarantees showing that policy improvements in the SM result in improvements in the original game, and we provide theoretical and empirical insights showing that the empirical variance can be controlled. Finally, we demonstrate a substantial improvement to previous methods in performance and computational speed via multiple experiments.

\section{Related Work}

The theory of learning in games largely considers complete-information finite games, hence, traditional techniques rely on discretization.
However, it is unclear how well a discretized strategy performs in the original continuous game in general \cite{waugh2009abstraction} and it suffers from the curse of dimensionality.
The first attempts to compute equilibria in imperfect-information auction games followed such an approach \cite{athey2001single} or expressed the game as a limit of a sequence of complete-information games \cite{armantier2008approximation}.  
In larger combinatorial auctions equilibria were first computed with an algorithm that computes pointwise best responses in a discretization of the strategy space via Monte Carlo integration \cite{bosshardComputingBayesNashEquilibria2020}. 
Besides the aforementioned NPGA, an approach that similarly learns continuous-action strategies was proposed \cite{li2021evolution}.
Both algorithms learn bid functions via zeroth-order gradient estimates that are used during simultaneous gradient ascent in self-play.
Our method considers a continuous surrogate game and enables the use of first-order gradient methods.

The idea of analytically smoothing markets is conceptually similar to that of differentiable physics simulations. Smooth approximations of the underlying dynamics were used in these simulations \cite{huang2021plasticinelab}.
Zeroth- and first-order methods were compared and the pros and cons of both when available were discussed \cite{suh2022differentiable}. Furthermore, they demonstrated that the presence of discontinuities in the objective causes the first-order estimator to be biased, whereas the zeroth-order estimator remains unbiased.
Smooth markets transfer these ideas to auctions.

\section{Preliminaries}
We restrict the formulations to the case of single-item auctions for brevity in the presentation. The extension to \emph{auctions of multiple independent items} is straightforward and we present some experimental results for both cases.

\subsection{Auctions as Bayesian Games}
A \emph{Bayesian auction game} is defined as a quintuple $G = (\mathcal I, \mathcal A, \mathcal V, F, u)$. $\mathcal I = \{1, \dots, n\}$ describes the set of bidders participating in the game. 
The set of possible bid profiles is given as $\mathcal A = \mathcal A_1 \times {\cdots} \times \mathcal A_{n}$, where $\mathcal A_i$ is the set of bids available to agent $i \in \mathcal I$. 
Whereas $\mathcal V = {\mathcal V_1 \times {\cdots} \times \mathcal V_{n}}$ is the set of \emph{valuation profiles}.
$F{:}\ \mathcal V \rightarrow [0, 1]$ defines the joint prior probability distribution over valuation profiles, which is assumed to be common knowledge among all agents and atomless. $F_i$ denotes agent $i$'s marginal distribution of valuations. In this study, the index $\text{-}i$ denotes a profile of valuations, bids, or strategies for all bidders, except bidder~$i$.

At the beginning of the game, nature draws a valuation profile $v \sim F$, and each agent $i$ is informed of his or her valuation $v_i \in \mathcal V_i$. We denote by $F_{i}$ the marginal distribution of bidder~$i$ and by $F_{\text{-}i|i}$ the conditional distribution of the opponents given $v_i$. Based on the drawn valuation $v_i$, each agent submits a bid $b_i$ according to the \emph{strategy}, \emph{policy}, or \emph{bid function} $\beta_i{:}\ \mathcal V_i \rightarrow \mathcal A_i$. We denote the resulting strategy space of bidder~$i$ as $\Sigma_i \subseteq \mathcal A_i^{\mathcal V_i}$ and the space of possible joint strategies as $\Sigma = \prod_i \Sigma_i$.

As part of the environment, the auctioneer collects these bids and applies an \emph{auction mechanism} that determines allocations $x_i \in \{0, 1\}$ for each bidder~$i$, such that the item is allocated to at most one bidder. Also, it determines payments $p(b) \in \mathbb{R}_{\geq 0}^n$ according to a payment rule $p$, which the agents must pay to the auctioneer. We will consider bidders with risk-neutral utility functions given by $u_i{:}\ \mathcal V_i \times \mathcal A \rightarrow \mathbb{R}$,
\begin{align}
	u_i(v_i, b) &= v_i \, x_i(b) - p_i(b) \label{eq:general-post-utility}\\
	&= 
	\begin{cases}
		v_i - p_i(b) & b_i > \max b_{\text{-}i},\\
		0 & \text{else},
	\end{cases}
	\label{eq:utility}
\end{align}
i.e., the players' utility is given by how much they value the good allocated to them minus the price to be paid. We will also write $u_i(v_i, b_i, b_{\text{-} i}) = u_i(v_i, b)$ with a slight abuse of notation. Thus, the bidders' utilities depend on all bidders' actions but only on their own valuations. They aim to maximize their utility $u_i$.
We omit bidders with risk aversion or other forms of utility and valuation correlations for brevity. Notwithstanding, our treatment of equilibrium computation also extends to these settings.
We will differentiate between the \emph{ex ante} state of the game, where bidders know only the prior $F$, the \emph{interim} state, where bidders additionally know their valuation $v_i \sim F_i$, and the \emph{ex post} state, where all bids have been submitted; thus, $u_i(v_i, b)$ can be evaluated.

\subsection{Equilibria}

\emph{Nash equilibria} (NE) are often regarded as the central solution concept in game theory. Informally, given the equilibrium strategy of the opponents in an NE, no agent has an incentive to unilaterally deviate.
\emph{Bayesian Nash equilibria} (BNE) extend this concept to games of incomplete information. Here, the expected utility over the distribution of opponent valuations is calculated instead.
For a private valuation $v_i \in \mathcal V_i$, bid $b_i \in \mathcal A_i$, and opponent strategies $\beta_{\text{-}i} \in \Sigma_{\text{-}i}$, we denote the \emph{interim utility} of bidder~$i$ as 
\begin{equation}
	\label{eq:iterim-util}
	\overline{u}_i(v_i, b_i, \beta_{\text{-}i}) \; = \; \mathbb{E}_{v_{\text{-}i}|v_i} [u_i(v_i, b_i , \beta_{\text{-}i}(v_{\text{-}i}))],
\end{equation}
where $v_{\text{-}i}|v_i$ denotes the expectation over the opponent's conditional prior distribution given the valuation $v_i$.
We also denote the \emph{interim utility loss} of bid $b_i$ incurred by not playing a best response, given $v_i$ and $\beta_{\text{-}i}$ by:
\begin{equation}
	\label{eq:ex-interim-loss}
	\overline \ell_i(v_i, b_i, \beta_{\text{-}i}) \; = \; \sup_{b'_i \in \mathcal A_i} \overline u_i(v_i, b'_i, \beta_{\text{-}i}) - \overline u_i(v_i, b_i, \beta_{\text{-}i}).
\end{equation}
An \emph{$\varepsilon$-Bayes Nash equilibrium} ($\varepsilon$-BNE) with $\varepsilon \geq 0$ is a strategy profile $\beta^* = (\beta^*_1, \dots, \beta^*_n) \in \Sigma$, such that no bidder can improve his or her interim expected utility more than $\varepsilon$ by deviating. Therefore, in an $\varepsilon$-BNE for all $i \in \mathcal I$, it holds that
\begin{equation}
	\label{eq:BNE}
	\sup_{v_i \in \mathcal{V}_i} \overline{\ell}_i(v_i, \beta_i^*(v_i), \beta^*_{\text{-}i}) \; \leq \; \varepsilon.
\end{equation}
A $0$-BNE is simply called a BNE. In a BNE, every bidder's strategy maximizes his or her expected interim utility across his or her valuation space, given the opponents' strategies. While BNEs are often defined at the \emph{interim} stage of the game, we also consider \emph{ex ante} equilibria as strategy profiles that concurrently maximize each bidder's \emph{ex ante} utility
\begin{equation}
	\tilde{u}_i(\beta_i, \beta_{\text{-}i}) \; = \; \mathbb{E}_{v_i} [\overline{u}_i(v_i, \beta_i(v_i), \beta_{\text{-}i})].
\end{equation}
To estimate the worst-case interim utility loss $\overline{\ell}_{\text{max}}$, we choose an equidistant grid of $n_\text{grid}$ alternative actions ranging from zero to the maximum valuation for all dimensions and calculate approximate best responses based on the average utility over a sample of $n_\text{batch}$ prior distributions. Taking the maximum over all valuations and bidders then gives an estimate of $\overline{\ell}_{\text{max}}$, bounding $\varepsilon$ for the ex ante case from above.

As a second metric, we additionally report the probability-weighted root mean squared error of the learned strategy $\beta_i$ to the exact BNE strategy $\beta^*_i$ for those settings where an analytical BNE is known. For a sample from the prior valuation of size $n_\text{batch}$, this approximates the $L_2$ distance $\lVert \beta_i - \beta_i^*\rVert_{\Sigma_i}$ of these two functions as
\begin{equation}
	\label{eq:L_2}
	L_2(\beta_i) \; = \; \left(\frac{1}{n_\text{batch}}\sum_{v_i} \left(\beta_i(v_{i}) - \beta^*_i(v_{i})\right)^2\right)^{1/2}.
\end{equation}
Unlike $\overline{\ell}_{\text{max}}$, this metric is much easier to compute and does not suffer the drawback that a strategy with a negatable small loss may still be arbitrarily distant from the actual BNE. However, it is only computable when an analytical BNE is available and may need multiple evaluations when there are multiple BNE.

\subsection{Gradient Optimization Methods}
Policy gradient methods are concerned with learning a parameterized policy $\beta_{\theta_i}$ that selects actions based on the current observations \cite{sutton2018reinforcement}. To maximize utility, bidder~$i$ updates the parameters $\theta_i$ according to gradient ascent.
This process is intended to compute approximate ex ante BNEs, that is, to find mutual best responses of the bidders for all possible valuations.
The exact gradient update for valuation $v_i$ in iteration $t$ is
\begin{align}
	\theta_i^t \; = \; \theta_i^{t-1} + \eta \cdot \nabla_{\theta_i^{t-1}} \, \overline{u}_i(v_i, \beta_{\theta_i^{t-1}}(v_i), \beta_{\theta_{\text{-} i}^{t-1}}).
\end{align}
This must be approximated in practice.
Two common methods are zeroth- and first-order gradient approximations. The former solely relies on evaluating the objective function $u_i$, whereas the gradient $\nabla_{\theta_i} u_i$ can be evaluated in the latter.

As stated in the introduction, the discontinuous nature of the ex post utility function stems from the sampling of the opponents' priors and their corresponding actions. We encounter $u_i$ from Equation~\ref{eq:utility} and its derivative (in general) is discontinuous in $b_i$. The observation of this inapplicability persists for all pricing regimes and behavioral assumptions that are commonly considered in auctions.
Thus, an unbiased gradient estimate of the interim utility function \emph{cannot} be derived by sampling the ex post gradient. Specifically, interchanging taking an expectation and differentiating is invalid: 
\begin{equation}
	\label{eq:operator_interchange}
	\nabla_{\theta_i} \, \mathbb{E}_{v_{\text{-} i}|v_i} [u_i] \; \neq \; \mathbb{E}_{v_{\text{-} i}|v_i} [\nabla_{\theta_i} \, u_i].
\end{equation}
We supply the mathematical details in Appendix~\ref{sec:leibniz}.
Therefore, the naive application of backpropagating the accumulated exact ex post gradients may not be expected to provide a meaningful estimate of the ex ante gradient. This study establishes a path towards valid first-order gradient estimates in auction games.

\subsection{Zeroth-Order Approximation Methods}
\cite{bichler2021learning} employed ES to circumvent the interchange of differentiation and integration. ES rely on a randomized finite difference approximation of the gradient based on perturbations in the parameter space of the neural networks which can be computed after averaging over the priors \cite{salimans2017evolution}.
This is an alternative zeroth-order method to the REINFORCE algorithm. Unlike ES, REINFORCE relies on perturbations in the action space by using mixed strategies (typically Gaussian distributions) such that the gradient of the action probability density can be approximated. \cite{salimans2017evolution} compared these estimates for RL applications and argued that the variance of the ES estimate can be significantly lower.
We overload the notation for the ease of readability and write $u_i(\theta_i, v_{\text{-} i}) = u_i(v_i, \beta_{\theta_i}(v_i), \beta_{\theta_{\text{-} i}}(v_{\text{-} i}))$.
For a hyperparameter $\sigma > 0$, the ES estimator can be derived from
\begin{align}
	& \nabla_{\theta_i} \, \mathbb{E}_{v_{\text{-} i}|v_i} \left[u_i(\theta_i, v_{\text{-} i})\right]\nonumber\\
	&\quad \approx \nabla_{\theta_i} \, \mathbb{E}_{\epsilon \sim \mathcal{N}(0, I)} \mathbb{E}_{v_{\text{-} i} |v_i} \left[u_i(\theta_i + \sigma \epsilon, v_{\text{-} i})\right]\\
	&\quad = \mathbb{E}_{\epsilon \sim \mathcal{N}(0, I)} \mathbb{E}_{v_{\text{-} i}|v_i} \left[\frac{\epsilon}{\sigma} u_i(\theta_i + \sigma \epsilon, v_{\text{-} i})\right].
\end{align}
The last term can now be approximated via sampling.
However, the ES gradient estimate comes at massive computational costs. It requires a large number of additional environment evaluations for the sampled population values of $\epsilon$. Parallelization is essentially unavailable, because it would reduce the number of samples from the prior when considering a fixed amount of memory. Latter of which is the main limiting factor in getting precise estimates of the expected utility in auction games. Thus, \cite{bichler2021learning} kept a large batch size and computed the ES sequentially using a default population size of 64.
Based on the variance of the estimate, \cite{salimans2017evolution} argued that ES are an attractive choice if the number of episodes is large, which is not the case for single-round auctions.


\section{Smoothing Single-Item Auctions}\label{sec:introduce-smoothing}
This section proposes the market-specific approach.

\subsection{Allocation and Price Smoothing}
The allocation of indivisible objects in auction games is typically modeled as a binary vector, with a one indicating that the item is allocated to the corresponding buyer. The set of legitimate allocations is defined as
\begin{equation}
	\mathcal{X} = \Bigl\{ x \in \{0, 1\}^{n} \,\Big|\,  \sum_{i=1}^n x_{i} \leq 1 \Bigr\}.
\end{equation}
For all commonly considered auctions, the allocations label the bids as winning or losing to maximize the auctioneer's revenue. They are calculated according to
\begin{equation}
	x(b) = \argmax_{x' \,\in\, \mathcal{X}} \sum_{i=1}^n b_{i} x'_{i}.
\end{equation}
Typical auction mechanisms only differ in their payment rules. Two noteworthy examples are the first-price mechanism, where bidders pay what they bid and the celebrated VCG mechanism (second-price), where they pay for the harm they cause others by competing \cite{krishna2009auction}.

These allocations result in the utilities not being continuous.
Therefore, we propose relaxing the calculation of the allocations using the softmax function as a surrogate for the argmax operation:
\begin{equation}
	x^{\text{SM}(\lambda)}_i(b) = \frac{\exp \left(\frac{b_{i}}{\lambda}\right)}{\sum_{j = 1}^n \exp \left(\frac{b_{j}}{\lambda}\right)}, \quad i = 1,\dots n.
\end{equation}
The temperature $\lambda > 0$ denotes the smoothing strength. This can be interpreted as dividing the item among all bidders according to their proportional bid magnitudes, where $\sum_{i} x^{\text{SM}(\lambda)}_{i}(b) = 1$ remains valid. The softmax asymptotically recovers the true argmax as $\lambda$ approaches zero.
As we are interested in a continuous utility surface, the discontinuity in the prices (only the winners pay) must also be considered. An obvious choice is to calculate the original prices of the good and then distribute the price according to the fractional allocations $x^{\text{SM}(\lambda)}$:
\begin{equation}
	p^{\text{SM}}(b) = \sum_{j=1}^n p_j(b).
\end{equation}
Hence, the ex post utility in the relaxed game takes the form
\begin{align}
	\label{eq:post-utility-smooth}
	u_i^{\text{SM}(\lambda)}(v_i, b) = \left(v_i - p^{\text{SM}}(b)\right) x_i^{\text{SM}(\lambda)}(b).
\end{align}
By definition, we have almost everywhere (a.e.) pointwise convergence of $x_i^{\text{SM}(\lambda)}(v_i, b_i, \beta_{\text{-}i}(\argdot))$ to $x_i(v_i, b_i, \beta_{\text{-}i}(\argdot))$ as functions of $v_{\text{-}i}$, except at $b_i = \max b_{\text{-}i}$. Furthermore, the fractional prices $p^{\text{SM}(\lambda)}(b_i, \beta_{\text{-}i}(\argdot))$ also converge a.e.\ pointwise to $p_i(b_i, \beta_{\text{-}i}(\argdot))$. Thus, the ex post utilities are recovered (a.e.) for ever smaller temperature. The resulting utilities are visualized for the special case of an FPSB auction (Figure~\ref{fig:ex-post-utility}).
Throughout the rest of the article, we make the following regularity assumptions.

\begin{assumption}
	\label{ass:main-assumptions-main-text}
	Consider a Bayesian auction game $G$ and assume:
	\begin{enumerate}
		\item The action $\mathcal A_i$ and valuation spaces $\mathcal V_i$ are compact intervals. \label{ass:regularities-interim-bound-val-action-space}
		\item $F$ is an atomless prior. \label{ass:regularities-interim-bound-atomless_prior}
		\item The bidding and pricing functions are measurable.
	\end{enumerate}
\end{assumption}

We regain continuity of the ex post utility and its gradient by this smoothing of allocations and payments. Specifically, we have the following theorem:
\begin{theorem} \label{thm:smoothing-gives-unbiased-gradient-estimator}
	Let the conditions of Assumption~\ref{ass:main-assumptions-main-text} hold and assume the pricing function $p^{\text{SM}}$, the marginal density functions $\{f_{\text{-}i|i}\}_{v_i \in \mathcal{V}_i, i \in \mathcal{I}}$, and strategies $\{\beta_i\}_{i \in \mathcal{I}}$ to be Lipschitz continuous. Then, the estimator on the smooth interim utility's gradient by sampling from the smoothed ex post utilities' gradients is unbiased, i.e.,
	\begin{align}
		\nabla_{\theta_i} \, \overline{u}^{\text{SM}}_i(v_i, b_i)
		= \mathbb{E}_{v_{\text{-} i}|v_i} [\nabla_{\theta_i} u^{\text{SM}}_i(v_i, b_i, \beta_{\text{-}i}(v_{\text{-}i}))],
	\end{align}
	for all $i \in \mathcal{I}$, $v_i \in  \mathcal{V}_i$, and $b_i \in \mathcal{A}_i$.
\end{theorem}

We refer to Appendix~\ref{sec:leibniz} for the proof.
Importantly, this relaxation technique is applicable to general markets with different payment rules, utility functions, or correlated priors.
Compared with the ES gradient estimate, where the parameter space is perturbed, the SM gradient estimate perturbs the utility function. Thus, the origin of bias is different and can be controlled by $\sigma$ for ES and by $\lambda$ for SM.

\subsection{Approximation Quality} \label{subsec:approximation-quality}
We check the validity of the smoothing intervention by ensuring that the error to the original game dynamics can be controlled by choosing a sufficiently small value of $\lambda$. This ensures that conducting policy optimization in the smoothed game can be expected to result in policy improvements in the original game. Furthermore, this will clarify the question of an optimal choice of the temperature value. 

Generally, analytically computing equilibria of the SM game is infeasible. Instead, we focus on comparing the expected interim and ex ante utilities in the original and SM game. A small error implies similar utility surfaces and gradient dynamics. 
Note that the ex post utilities can be quite different. Suppose multiple bidders compete for a single commodity and bidder~$i$ has approximately the same bid magnitude as the strongest opponent. The smoothed allocation is close to one-half, whereas the true allocation is either zero or one. This would result in a significant difference in the ex post utility driven by the magnitude of the utility discontinuity in the original auction. The probability of such large errors decreases with smaller smoothing factors; however, this event cannot be completely ruled out.
We verify in the following theorem, that the error in expected interim and ex ante utility approaches zero under mild assumptions on the auction format.

\begin{theorem}\label{thm:asymptotic}
	Let the conditions of Assumption~\ref{ass:main-assumptions-main-text} hold and suppose the payment rule $p$ is bounded. Then, for bidder~$i$, we have convergence in interim and ex ante utility: 
	\begin{enumerate}
		\item Let $v_i \in \mathcal V_i$ and $b_i \in \mathcal A_i$, then
		\begin{equation}
			\lim_{\lambda \rightarrow 0} \overline{u}_i^{\text{SM}(\lambda)}(v_i, b_i, \beta_{\text{-}i}) \; = \; \overline{u}_i(v_i, b_i, \beta_{\text{-}i}).
		\end{equation}
		\item Further assume $\beta_i$ to be measurable. Then,
		\begin{equation}
			\lim_{\lambda \rightarrow 0} \tilde{u}_i^{\text{SM}(\lambda)}(\beta_i, \beta_{\text{-}i}) \; = \; \tilde{u}_i(\beta_i, \beta_{\text{-}i}).
		\end{equation}
	\end{enumerate}
\end{theorem}
The proof is delegated to Appendix~\ref{sec:proof-asymptotic}.
Theorem~\ref{thm:asymptotic} ensures that for ever smaller $\lambda$, the bias in the expected utilities vanishes compared with the utilities in the original game. This implies that the smoothed gradients converge, thus justifying gradient-based learning in the perturbed game. 
Although Theorem~\ref{thm:asymptotic} ensures convergence, it does not state how fast the error approaches zero. However, this information is crucial for practical applications. Therefore, we make the following additional assumptions on the auction format.

\begin{assumption} \label{ass:regularities-for-interim-bound}
	For all $i \in \mathcal{I}$ assume:
	\begin{enumerate}
		\item $\beta_i$ is strictly increasing and Lipschitz continuous. \label{ass:regularities-interim-bound-strategies}
		\item $\beta_i^{-1}$ is Lipschitz continuous. \label{ass:regularities-interim-bound-inverse}
		\item There exists a uniform bound for all marginal conditional prior density functions $f_{i|\argdot}$. \label{ass:regularities-interim-bound-prior}
		\item $p_i$ is bounded. \label{ass:regularities-interim-bound-pricing}
	\end{enumerate}
\end{assumption}

Note that assuming Lipschitz continuous strategies is satisfied by common function approximations, e.g., neural networks.
With these stronger assumptions, we can present a worst-case convergence rate of the interim and ex ante utility errors. 

\begin{proposition} \label{thm:linear-convergence-utility-error}
	Consider an auction with $n$ bidders that satisfies Assumptions~\ref{ass:main-assumptions-main-text} and \ref{ass:regularities-for-interim-bound}. Then, the absolute interim and ex ante utility errors are of order $\mathcal{O}(\lambda)$. 
\end{proposition}

\begin{proof}[Proof Sketch]
	Use substitution on the opponents' bidding strategies, followed by iterated use of Hölder's inequality. The details of the proof can be found in Appendix~\ref{sec:proof-linear-conv}.
\end{proof}

Note that Restrictions~\ref{ass:regularities-interim-bound-strategies} and \ref{ass:regularities-interim-bound-pricing} in Assumption \ref{ass:regularities-for-interim-bound} are standard in the literature \cite{krishna2009auction}. Restriction \ref{ass:regularities-interim-bound-inverse} is slightly stronger by demanding that strategy $\beta_i$ cannot become infinitely flat (e.g., a saddle-point would not be allowed). However, this restriction can be somewhat lifted resulting in a worse convergence rate. Details on this can be found in Appendix~\ref{sec:proof-linear-conv}. Finally, Restriction~\ref{ass:regularities-interim-bound-prior} holds for all commonly used prior distributions, however, it rules out perfect correlation.
Based on the previous result, we can characterize how a learned $\varepsilon$-BNE of the SM game translates to an approximate BNE the original game:

\begin{theorem} \label{thm:eps-equ-in-smooth-game-to-original-game}
	In an auction with $n$ bidders that satisfies Assumptions~\ref{ass:main-assumptions-main-text} and \ref{ass:regularities-for-interim-bound}, let $\beta^*$ be an ex ante $\varepsilon$-BNE in the smoothed game with smoothing parameter $\lambda$. Then $\beta^*$ is an ex ante $\varepsilon + \mathcal{O}(\lambda)$-BNE of the original game. 
\end{theorem}

\begin{figure}[t]
	\centering
	\includegraphics[width=0.85\columnwidth]{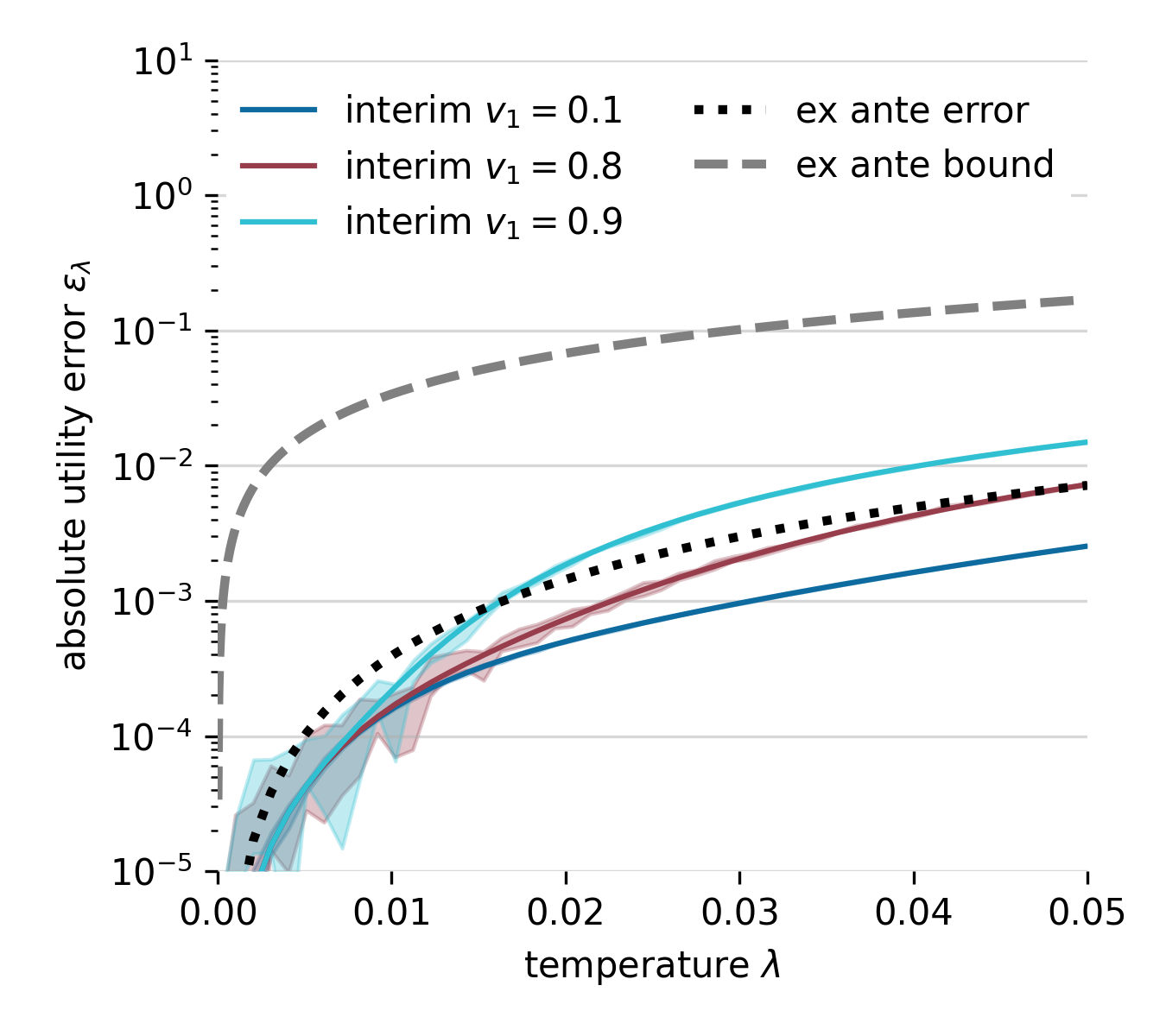}
	\caption{Comparison of the absolute utility errors. (i) The linear ex ante bound that holds for all valuations (gray dashed line) from Equation~\ref{eq:linear-ante-bound}. (ii) Exact interim utility errors when both bidders act according to the BNE for some exemplary valuations (colorized lines) and their sampled mean values $\pm$ standard deviation (shaded areas). (iii) The approximate ex ante error (black dotted line).}
	\label{fig:smooth_error_bound}
\end{figure}

The proof can be found in Appendix~\ref{sec:proof-cor-1}. 
The derived bounds in the previous results consider worst-case scenarios. However, we observed that the error may be significantly lower in practice. To rationalize this observation, we compare the worst-case bound to the exact error in a restricted setting.
Consider an FPSB auction with two bidders, independent uniform priors, and a linear bidding function of the second bidder, $\beta_2(v_2)=sv_2 + t$. Then, the bound derived in Proposition~\ref{thm:linear-convergence-utility-error} translates to
\begin{align}\label{eq:linear-ante-bound}
	\left| \tilde{u}_1^{\text{SM}(\lambda)}(\beta_1, \beta_2) - \tilde{u}_1(\beta_1, \beta_2) \right| \; \leq \; \frac{\ln(2) + 1}{s} \lambda.
\end{align}
In Figure~\ref{fig:smooth_error_bound}, we compare this bound (for bidder~2's BNE strategy with $s=0.5$ and $t=0$) to the exact interim utility error, which can be derived for this restricted setting (see Appendix~\ref{sec:exact-interim-error}).
The convergence rate of the interim utilities depends on the specific prior sample $v_1$ and bid $b_1$. The ex ante utilities converge more rapidly than predicted by the worst-case bound. We conjecture that this often holds in practice, resulting in better learning behavior than suggested by Proposition~\ref{thm:linear-convergence-utility-error}.

\subsection{Choosing the Smoothing Temperature}\label{sec:optimal-temp}
Let us consider the question of an optimal smoothing strength.
There is an incentive to keep temperature values as low as possible, such that the original game dynamics are distorted as little as possible. On the other hand, one does not want to decrease $\lambda$ too low, as this causes numerical problems. The magnitude of the gradient goes towards infinity at the former discontinuity as $\lambda$ decreases. Therefore, with finite sample size, the first-order gradient estimate might have a high empirical variance \cite{suh2022differentiable}.

We propose to use the utility sampling precision as a natural way to choose the temperature. For the special case presented in Figure~\ref{fig:smooth_error_bound} and the default batch size of $2^{18}$, one can see that the sample precision is reached at about $10^{\text{-}4}$. That is, for a drawn batch, the Monte Carlo estimation of ex ante utilities has a precision of about $10^{\text{-}4}$, and we can no longer distinguish between the smoothed and original utilities. Therefore, one can use Proposition~\ref{thm:linear-convergence-utility-error} to derive a lower bound for $\lambda$ for a given sampling precision. As discussed at the end of Section~\ref{subsec:approximation-quality}, the true ex ante utility error is usually lower, so that one can choose a higher $\lambda$ without losing any performance. 

The empirical sampling precision is affected by several factors, such as the valuation and bidding ranges, the number of bidders, prior distributions, and complexity of bidding functions. Some of these influences can be standardized, e.g., by normalizing the bidding ranges. Ultimately, a sufficiently high batch size can overcome any bias introduced by aforementioned factors, such that it should be chosen as high as computationally possible to achieve an optimal sampling precision.

\section{Empirical Results}\label{sec:results}

We provide experimental evaluation of the new technique and compare the results with those of NPGA and REINFORCE by measuring how closely they approximate the analytical BNE. Results for settings with risk aversion or correlated valuations are similar and omitted for simplicity.
Furthermore, we provide some insights and guidance on appropriate choices of $\lambda$ and verify that our gradient estimate's variance is sufficiently small.
We list all hyperparameters and details on the network architecture in Appendix~\ref{sec:reproducibility}.

\begin{table}[t]\fontsize{8}{10}\selectfont
	\centering
	\caption{Learning results in FPSB and SPSB auctions with different numbers $m$ of items. We report the mean values of the $L_2$ and $\overline{\ell}_{\text{max}}$ losses (smaller is better) and the time per iteration across five runs. We also report the standard deviation in parentheses for the losses.}
	\vspace{0.1in}  
	\begin{tabular}{lllrrr}
		\toprule
		        			  & $m$ 				   & Algorithm 	   & $L_2$ 			& $\overline{\ell}_{\text{max}}$ & $t/$iter \\
		\midrule
		\parbox[t]{2mm}{\multirow{12}{*}{\rotatebox[origin=c]{90}{FPSB}}} & \multirow{3}{*}{1} & NPGA	   &  0.011 (0.005) &       0.005 (0.002) &  0.155 \\
																		&					   & REINFORCE &  0.021 (0.008) &       \textbf{0.003 (0.000)} &  \textbf{0.009} \\
																		&					   & SM 	   &  \textbf{0.005 (0.003)} &       0.004 (0.002) &  \textbf{0.009} \\
																		\cline{2-6}
																		& \multirow{3}{*}{2}   & NPGA      &  0.013 (0.005) &       0.010 (0.002) &  0.150 \\
																		&					   & REINFORCE &  0.041 (0.020) &       0.016 (0.010) &  \textbf{0.009} \\
																		&					   & SM 	   &  \textbf{0.008 (0.002)} &       \textbf{0.006 (0.003)} &  \textbf{0.009} \\
																		\cline{2-6}
																		& \multirow{3}{*}{4}   & NPGA      &  0.028 (0.002) &       0.021 (0.003) &  0.148 \\
																		&   				   & REINFORCE &  0.064 (0.018) &       0.039 (0.012) &  \textbf{0.009} \\
																		&   				   & SM 	   &  \textbf{0.015 (0.004)} &       \textbf{0.011 (0.004)} &  \textbf{0.009} \\
																		\cline{2-6}
																		& \multirow{3}{*}{8}   & NPGA      &  0.104 (0.054) &       0.127 (0.109) & 0.206 \\
																		&   				   & REINFORCE &  0.187 (0.073) &       0.331 (0.169) & \textbf{0.012} \\
																		&   				   & SM 	   &  \textbf{0.036 (0.003)} &       \textbf{0.034 (0.009)} & \textbf{0.012} \\
		\cline{1-6}
		\parbox[t]{2mm}{\multirow{12}{*}{\rotatebox[origin=c]{90}{SPSB}}} & \multirow{3}{*}{1} & NPGA      &  0.012 (0.001) &       0.002 (0.000) &  0.170 \\
																		&   				   & REINFORCE &  0.028 (0.005) &       0.002 (0.000) &  \textbf{0.009} \\
																		&   				   & SM 	   &  \textbf{0.004 (0.001)} &       \textbf{0.001 (0.000)} &  0.011 \\
																		\cline{2-6}
																		& \multirow{3}{*}{2}   & NPGA      &  0.018 (0.002) &       0.003 (0.000) &  0.264 \\
																		&   				   & REINFORCE &  0.082 (0.020) &       0.009 (0.002) &  0.011 \\
																		&   				   & SM 	   &  \textbf{0.007 (0.001)} &       \textbf{0.002 (0.000)} &  \textbf{0.015} \\
																		\cline{2-6}
																		& \multirow{3}{*}{4}   & NPGA      &  0.043 (0.002) &       0.011 (0.003) &  0.457 \\
																		&   			       & REINFORCE &  0.140 (0.045) &       0.028 (0.018) &  \textbf{0.017} \\
																		&   				   & SM	       &  \textbf{0.029 (0.003)} &       \textbf{0.006 (0.002)} &  0.024 \\
																		\cline{2-6}
																		& \multirow{3}{*}{8}   & NPGA      &  0.214 (0.112)	&       0.299 (0.238) &  0.869 \\
																		&   				   & REINFORCE &  0.320 (0.128) &       0.262 (0.174) &  \textbf{0.031} \\
																		&   				   & SM 	   &  \textbf{0.074 (0.002)} &       \textbf{0.020 (0.002)} &  0.043 \\
		\bottomrule
	\end{tabular}
	\label{tab:single_item}
\end{table}

\subsection{Single-Item Auctions}
For the two common payment rules of FPSB and second-price sealed-bid (SPSB) and a uniform prior on $[0,1]$, we can measure the distance in action space to the unique BNE, as described in Equation~\ref{eq:L_2} and compute an estimate of exploitability in the form of Equation~\ref{eq:BNE}. Table~\ref{tab:single_item} shows the results. The losses are computed after training 2,000 iterations with each algorithm. The time per iteration, $t$/iter, decreases notably when comparing NPGA to SM across both payment rules, while also achieving a better approximation quality. Since the estimation of $\overline{\ell}_{\text{max}}$ relies on a discretization of the action space and an exhaustive search thereon, $L_2$ detects smaller deviations, ceteris paribus. Although REINFORCE has a low iteration time, it is unable to learn high quality strategies due to its high variance (Section~\ref{sec:variance-empirical}).
We found that results for auctions with interdependent prior valuations or risk-aversion are quantitatively consistent with the results presented here.


\subsection{Large Simultaneous Auctions}
Furthermore, we study the separate sales of up to $m=8$ distinctive goods and an increase in the number of bidders of up to $n=4$.
For simplicity, we do not consider any synergy effects on the items (this would include cases such as those where a bidder only values the bundle of two items but not either one of them individually), such that the BNE simplifies to the single-item strategy profile for each item separately.
There are multiple motivations for these auctions. They can be considered as the base case of combinatorial auctions with item bidding and as a simple and practical alternative to full combinatorial auctions. Furthermore, combinatorial auctions with item bidding are being deployed, e.g., a bidder who is interested in a bundle of objects in parallel online display ad auctions or on a consumer shopping website is implicitly partaking in these auctions.
Finally, asking a bidder to submit bids on all possible combinations of bundles ($2^m - 1$) is practically infeasible and there are positive results on the welfare properties of limiting the action space in this way \cite{bhawalkar2011welfare}.
Again, we draw i.i.d.\ uniform valuations on $[0, 1]$ and consider the FPSB and SPSB auctions.
Learning in the SM game outperforms both previous approaches (Table~\ref{tab:single_item}).
Since first-order methods are generally faster, we assume that the strong results in these settings will scale to even larger ones.



\begin{figure}[t]
	\centering
	\includegraphics[width=1.0\columnwidth]{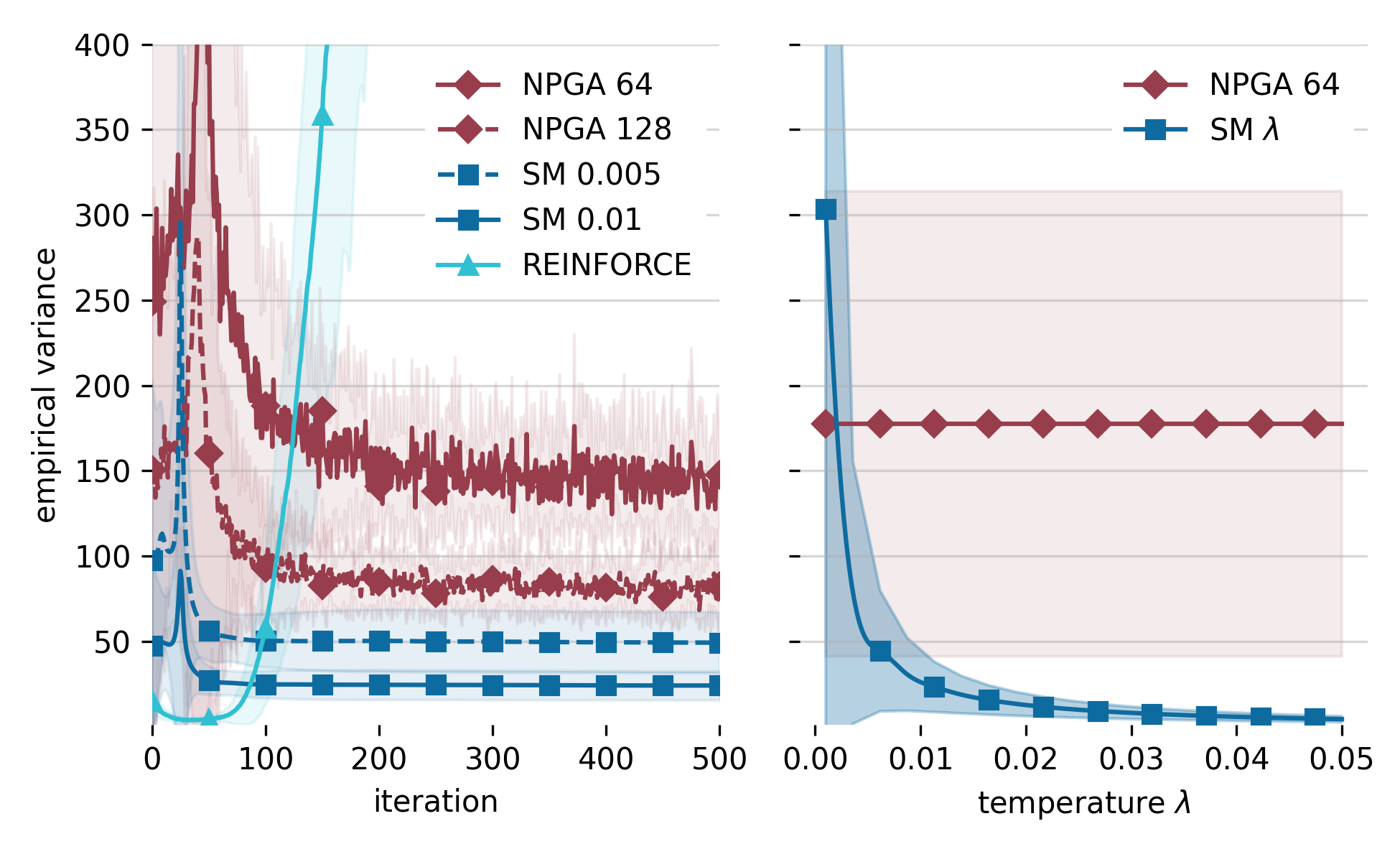}
	\caption{Empirical variance of the NPGA and REINFORCE zeroth-order and the SM first-order gradient estimates. Both zeroth-order methods are run in the original auction game. The mean values $\pm$ standard deviations over five runs each are depicted. \emph{Left:} Comparing the variance throughout the learning procedure. \emph{Right:} Comparing the variance for different smoothing temperatures (averaged over complete training runs).}
	\label{fig:variance_analysis}
\end{figure}

\begin{figure}[t]
	\centering
	\includegraphics[width=0.85\columnwidth]{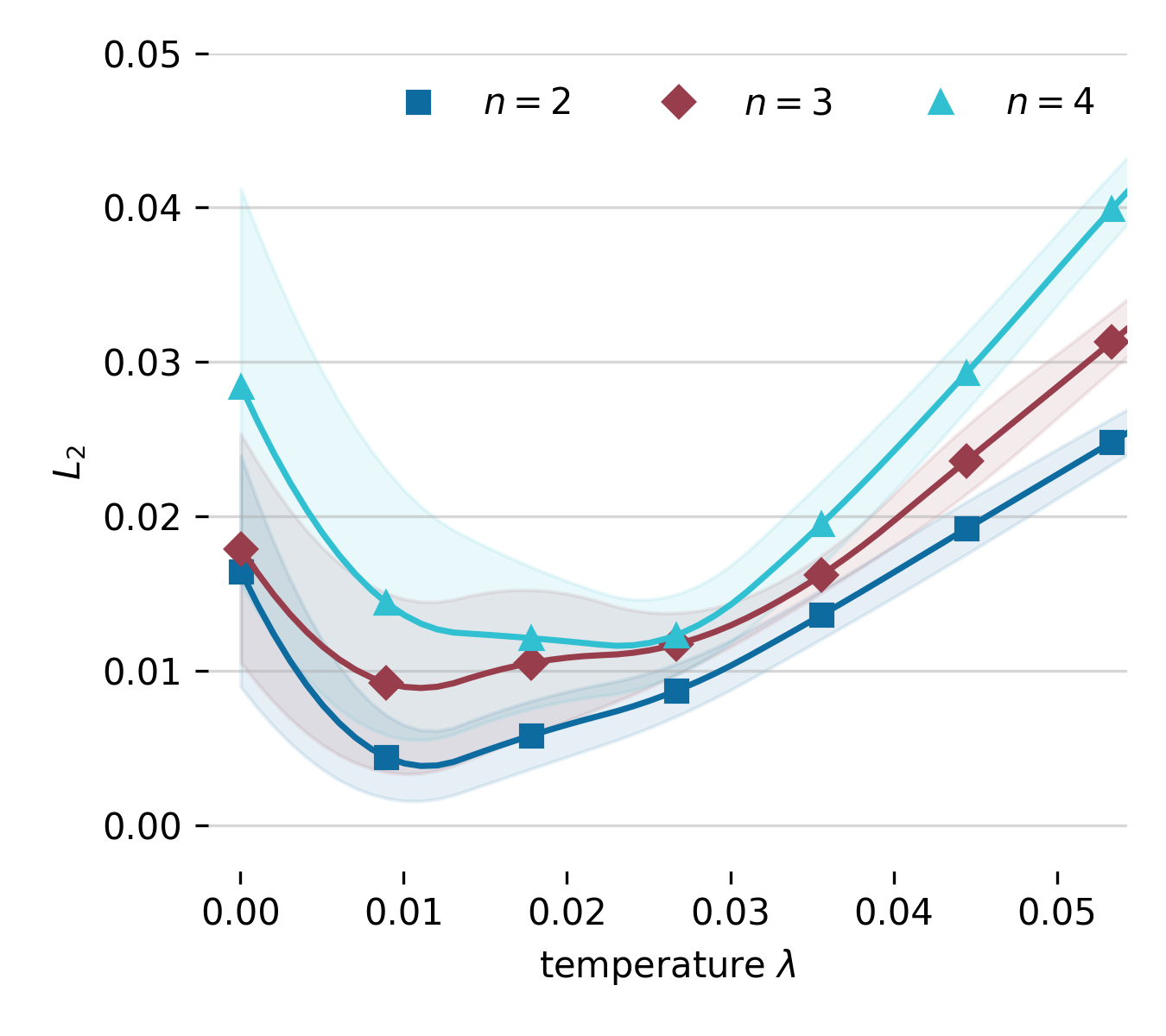}
	\caption{Action space distance for learned strategies to the BNE for different numbers of bidders and temperature values $\lambda$. The mean values $\pm$ standard deviations over five runs each are depicted.}
	\label{fig:temperature_analysis}
\end{figure}

\subsection{Empirical Variance}\label{sec:variance-empirical}
As stated in Section~\ref{sec:optimal-temp}, there is a trade-off between low and high values of $\lambda$. 
Here, we consider the base setting of two bidders competing in a single-item FPSB auction. We decrease the batch size to $2^{16}$ as the single-sample gradients require more memory.
Considering NPGA that is based on a sample of 64 evaluations of the objective by default, the empirical variance of the SM estimate is lower for all $\lambda > 0.002$ (compare intersection of Figure~\ref{fig:variance_analysis}, right plot). Even after increasing NPGA's population size by a factor of two (which scales the run time in the same way), SM's variance remains lower for most choices, as can be seen in the left figure.
The empirical variance of REINFORCE rapidly increases as the mixed-strategies get closer to the pure-strategy BNE. This degradation is to be expected when the learned variance of the Gaussian distributed actions decreases, see Exercise~13.4 of \cite{sutton2018reinforcement}.

Results for markets of different sizes are depicted in Figure~\ref{fig:temperature_analysis}. Keeping everything else fixed, the highest achievable performance decreases for larger markets, as is expected in multi-agent learning. The optimal smoothing strength is only affected indirectly via the bid magnitudes. 
At last, we note that the performance boost of larger batch sizes diminishes and best results are achieved for similar values of $\lambda$ just below $0.01$, indicating that the variance of the gradient estimate counteracts the lower bias. The results are presented in Appendix~\ref{sec:bacth-size}.

\section{Conclusion and Future Work}

How can first-order gradient estimation methods be successfully applied to learning in auctions?
We showed that our proposed smooth game formulation of strategic interactions in auctions provides a strong answer to this question. We established theoretical bounds on the bias caused by the smoothing, and an empirical evaluation verified that the variance of the gradient estimate can be controlled, leading to low computational costs and high precision. Overall, we verified that equilibrium computation in smooth markets via fist-order gradient estimation is more efficient than previous learning methods.


\input{ms.bbl}
\bibliographystyle{icml2023}

\newpage
\appendix
\onecolumn

\input{technical-appendix.tex}

\end{document}

%% file: technical-appendix.tex
\section{Proof of Theorem~\ref{thm:smoothing-gives-unbiased-gradient-estimator}}\label{sec:leibniz}

The Leibniz integral rule states conditions for which the operator interchange of taking the limit and integrating is valid. Let us recall its measure theory variant using the notation of our application. For a rigorous treatment of the assumptions and different variants, we refer to the work of \citet{talvilaNecessarySufficientConditions2001}, see Corollaries~5 and 8. We reformulate Condition~1 in our version using Fubini's Theorem.
\begin{theorem}[Leibniz integral rule]\label{thm:leibniz}
	Let $[a, b] \subset \mathbb{R}$ and $\Omega$ with $\mu$ be a probability space. Suppose $f: [a, b] \times \Omega \to \mathbb{R}$ satisfies the conditions: 
	\begin{enumerate}
		\item $f(x, \omega)$ is a measurable function in $x$ and $\omega$, and is integrable over $\omega \in \Omega$, for almost all $x \in [a, b]$. \label{thm:leibniz-condition-measurability-and-integrability}
		\item For almost all $\omega \in \Omega$, $f(x, \omega)$ is absolutely continuous in $x$. \label{thm:leibniz-condition-absolutely-continuous}
		\item For all compact intervals $[c, d]\subset [a, b]$:
		 \begin{align}
		 	\int_c^d \int_{\Omega} \left| \frac{\partial}{\partial x} f(x, \omega) \right| \mathrm{d} \mu(\omega) \mathrm{d} x < \infty.
		 \end{align} \label{thm:leibniz-condition-locally-integrable}
	\end{enumerate}
	Then 
	\begin{equation}
		\label{eq:leibniz}
		\frac{\partial}{\partial x} \int_{\Omega} f(x, \omega) \, \mathrm{d}\mu(\omega) \; = \; \int_{\Omega} \frac{\partial}{\partial x} f(x, \omega)\, \mathrm{d} \mu(\omega) \qquad \text{a.e.}
	\end{equation}
\end{theorem}

Equation~\ref{eq:leibniz} is assumed to hold for backpropagation. It is needed for the approximation of ex ante gradients based the sample mean of ex post gradients (compare Equation~\ref{eq:operator_interchange} from main text). We are now ready to prove Theorem~\ref{thm:smoothing-gives-unbiased-gradient-estimator}.

\begin{proof}
	We show that under the assumptions made in Theorem~\ref{thm:smoothing-gives-unbiased-gradient-estimator}, the Leibniz integral rule as formulated in Theorem~\ref{thm:leibniz} holds. We proceed to show that the smooth ex post utilities $u_i^{\text{SM}(\lambda)}$ are Lipschitz continuous, which essentially ensures all three conditions hold.
	So, for $i \in \mathcal{I}$, recall the smoothed ex post utility for some $v_i \in \mathcal{V}_i$ and $\lambda >0$:
	\begin{align}
		u_i^{\text{SM}(\lambda)}(v_i, b_i, \beta_{\text{-} i}(v_{\text{-}i})) = \left(v_i - p^{\text{SM}}(b_i, \beta_{\text{-} i}(v_{\text{-}i})) \right) x_i^{\text{SM}(\lambda)}(b_i, \beta_{\text{-} i}(v_{\text{-}i})).
	\end{align}
	As the smooth pricing function $p^{\text{SM}}$ is a sum over Lipschitz continuous functions, it is Lipschitz continuous. As $\beta_{\text{-} i}$ and $x_i^{\text{SM}(\lambda)}$ are Lipschitz continuous, so is $x_i^{\text{SM}(\lambda)}(b_i, \beta_{\text{-} i}(\argdot))$. Finally, as both $p^{\text{SM}}$ and $x_i^{\text{SM}(\lambda)}$ are bounded and Lipschitz continuous, their product is Lipschitz continuous as well. Therefore, $u_i^{\text{SM}(\lambda)}$ is Lipschitz continuous in $b_i$ and $v_{\text{-}i}$. 
	
	The Lipschitz continuity of $u_i^{\text{SM}}$ ensures measurability, as well as integrability over $\mathcal{V}_{\text{-}i}$ for all $b_i \in \mathcal{A}_i$. Hence, Condition~\ref{thm:leibniz-condition-measurability-and-integrability} holds. As Lipschitz continuity is stronger than absolute continuity, Condition~\ref{thm:leibniz-condition-absolutely-continuous} holds as well. Finally, note that due to Lipschitz continuity, there exists an $L$ such that $\left| \frac{\partial}{\partial b_i} u_i^{\text{SM}(\lambda)}(v_i, b_i, \beta_{\text{-} i}(v_{\text{-}i})) \right| \leq L$ for all $b_i \in \mathcal{A}_i$. This bound ensures that Condition~\ref{thm:leibniz-condition-locally-integrable} holds also. 
	
	In the case of conditional priors, consider the function $u_i^{\text{SM}(\lambda)} f_{\text{-}i|\argdot}$. This function is again Lipschitz continuous as product of bounded Lipschitz continuous functions. Repeating the steps above for this function finishes the proof. 
\end{proof}

\begin{remark}
	The original non-smooth ex post utility function $u_i$ does not satisfy the conditions of Theorem \ref{thm:leibniz}. For example, $u_i$ is not even continuous in $b_i$, so that the second condition is violated.
\end{remark}

\section{Proof of Theorem \ref{thm:asymptotic}}\label{sec:proof-asymptotic}

\begin{proof}
	Let us start with the first statement. For the interim utility of bidder~$i$ in the original game, we have
	\begin{align}
		\label{eq:iterim-utility}
		\overline{u}_i(v_i, b_i, \beta_{\text{-}i}) &= \mathbb{E}_{v_{\text{-}i}|v_i}\left[ u_i(v_i, b_i, \beta_{\text{-}i}(v_{\text{-}i}))\right]
	\end{align}
	with the ex post utility from Equation~\ref{eq:general-post-utility} in the main text rewritten as
	\begin{align}
		\label{eq:post-utility-original}
		u_i(v_i, b_i, b_{\text{-}i}) = (v_i - p_i(b_i, b_{\text{-}i})) x_i(b_i, b_{\text{-}i}).
	\end{align}
	In the smoothed auction, we have
	\begin{align}
		\label{eq:smooth-iterim-utility}
		\overline{u}_i^{\text{SM}(\lambda)}(v_i, b_i, \beta_{\text{-}i}) = \mathbb{E}_{v_{\text{-}i}|v_i}\left[ u_i^{\text{SM}(\lambda)}(v_i, b_i, \beta_{\text{-}i}(v_{\text{-}i})) \right]
	\end{align}
	with $u_i^{\text{SM}(\lambda)}$ as defined in Equation~\ref{eq:post-utility-smooth} in the main text.
	Note that $u_i^{\text{SM}(\lambda)}$ is integrable as composition of integrable functions.
	We first have a.e.\ pointwise convergence of $u_i^{\text{SM}(\lambda)}(v_i, b_i, \beta_{\text{-}i}(v_{\text{-}i}))$ to $u_i(v_i, b_i, \beta_{\text{-}i}(v_{\text{-}i}))$ as $\lambda$ approaches zero. That is, for all $v_{\text{-}i}$, except for $b_i = \beta_{\text{-}i}(v_{\text{-}i})$, the smaller $\lambda$ gets the closer the allocations and the closer the utilities get.

	Second, it is easy to see that $\lvert u_i^{\text{SM}(\lambda)}(v_i, b_i, \beta_{\text{-}i}(v_{\text{-}i})) \rvert$ is bounded via
	\begin{align*}
		&\lvert u_i^{\text{SM}(\lambda)}(v_i, b_i, \beta_{\text{-}i}(v_{\text{-}i})) \rvert \leq \lvert v_i - p^{\text{SM}}(b_i, \beta_{\text{-} i}(v_{\text{-}i})) \rvert
	\end{align*}
	and noting that this is a composition of bounded functions.
	With these two conditions satisfied, we can apply the dominated convergence theorem in its a.e.\ version (see, e.g., \citet{bogachev2007measure}, Theorem~2.8.1) on the terms from Equations \ref{eq:iterim-utility} and \ref{eq:smooth-iterim-utility} which proofs the first statement. 

	Let us now consider the ex ante utilities. From the interim convergence, we know that the expected interim utility $\overline{u}_i^{\text{SM}(\lambda)}(v_i, b_i, \beta_{\text{-}i})$ converges pointwise to $\overline{u}_i(v_i, b_i, \beta_{\text{-}i})$ for all $v_i$ and $b_i$. Again applying the dominated convergence theorem ensures equality of the expected utility in the ex ante state of the game. 
\end{proof}

\begin{remark}
	Technically, a tie-breaking rule should be specified for $x_i$ at the nullset $b_i = \beta_{\text{-}i}(v_{\text{-}i})$, but that is exactly the point which we neglect.
\end{remark}

\section{Proof of Proposition \ref{thm:linear-convergence-utility-error}}\label{sec:proof-linear-conv}
The following section provides a linear bound on the error in interim and ex ante utility. For clarity, we restate all major assumptions.

\begin{assumption}
	Consider a Bayesian auction game $G$ and assume:
	\begin{enumerate}
		\item The action $\mathcal A_i$ and valuation spaces $\mathcal V_i$ are compact intervals.
		\item $F$ is an atomless prior.
		\item The bidding and pricing functions are measurable.
	\end{enumerate}
\end{assumption}

\begin{assumption}
	For all $i \in \mathcal{I}$ assume:
	\begin{enumerate}
		\item $\beta_i$ is strictly increasing and Lipschitz continuous. 
		\item $\beta_i^{-1}$ is Lipschitz continuous.
		\item There exists a uniform bound for all marginal conditional prior density functions $f_{i|\argdot}$.
		\item $p_i$ is bounded.
	\end{enumerate}
\end{assumption}

\begin{proof}
	
We use the H\"older inequality throughout the proof, which we denote by (H). Whenever we use a specific assumption, we denote it by the corresponding number. 
We begin with the interim utility error, i.e., we aim to bound the following term for all $i \in \mathcal{I}, v_i \in \mathcal{V}_i$, and $\lambda >0$:
\begin{align}
	\overline{\varepsilon}_i(v_i, b_i, \beta_{\text{-} i}, \lambda) 
	:= \left| \int_{\mathcal{V}_{\text{-}i}} u_i\left(v_i, b_i, \beta_{\text{-} i}(v_{\text{-}i}) \right) - u_i^{\text{SM}(\lambda)}\left(v_i, b_i, \beta_{\text{-} i}(v_{\text{-}i}) \right) \mathrm{d} F_{\text{-}i|i}(v_{\text{-}i}) \right| .
\end{align}
So, let $i \in \mathcal{I}$, $v_i \in \mathcal{V}_i$, $b_i \in \mathcal{A}_i$, and $\lambda >0$ be arbitrary. 
By splitting up the integral into the individual opponents, we get
\begin{align}
	\overline{\varepsilon}_i(v_i, b_i, \beta_{\text{-} i}, \lambda)
	& = \Bigg\vert \int_{\mathcal{V}_{1}} \cdots \int_{\mathcal{V}_{i-1}} \int_{\mathcal{V}_{i+1}} \cdots \int_{\mathcal{V}_{n}} u_i\left(v_i, b_i, \beta_{\text{-} i}(v_{\text{-}i}) \right) - u_i^{\text{SM}(\lambda)}\left(v_i, b_i, \beta_{\text{-} i}(v_{\text{-}i}) \right) \nonumber\\
	& \qquad \mathrm{d} F_{n|\text{-}n}(v_{n}) \dots \mathrm{d} F_{i+1|(1, \dots, i)}(v_{i+1}) \mathrm{d} F_{i-1|(1, \dots, i-2, i)}(v_{i-1}) \dots \mathrm{d} F_{1|i}(v_{1}) \Bigg\vert\\
	\overset{(\text{H}), (\text{A}1.\ref{ass:regularities-interim-bound-atomless_prior}), (\text{A}2.\ref{ass:regularities-interim-bound-prior})}&{\leq}
	\prod_{j \neq i} \norm{f_j}_{\infty} \int_{\mathcal{V}_{\text{-}i}} \left| u_i\left(v_i, b_i, \beta_{\text{-} i}(v_{\text{-}i}) \right) - u_i^{\text{SM}(\lambda)}\left(v_i, b_i, \beta_{\text{-} i}(v_{\text{-}i}) \right)  \right| \mathrm{d} v_{\text{-}i} =: (*_1).
\end{align}
We use $\norm{f_j}_{\infty}$ to denote the uniform bound for any marginal conditional prior density function $f_j|\cdot$ of bidder $j$. Next, we perform a change of variables \citep{katzChangeVariablesMultiple1982} through the inverse of the opponents' strategies
\begin{align}
	(*_1) \overset{(\text{A}2.\ref{ass:regularities-interim-bound-strategies}), (\text{A}2.\ref{ass:regularities-interim-bound-inverse})}&{=}
	 \prod_{j \neq i} \norm{f_j}_{\infty} \int_{\beta_{\text{-} i}\left(\mathcal{V}_{\text{-}i}\right)} \left| \left( u_i\left(v_i, b_i, b_{\text{-}i} \right) - u_i^{\text{SM}(\lambda)}\left(v_i, b_i, b_{\text{-}i} \right) \right) \right| \cdot \left| \det \text{D}\beta_{\text{-} i}^{-1}(b_{\text{-}i}) \right| \mathrm{d} b_{\text{-}i} \\
	\overset{(\text{H}), (\text{A}1.\ref{ass:regularities-interim-bound-val-action-space})}&{\leq} \prod_{j \neq i} \norm{f_j}_{\infty} \cdot \norm{\left(\beta^{-1}_j\right)^{\prime}}_{\infty} \int_{\beta_{\text{-} i}\left(\mathcal{V}_{\text{-}i}\right)} \left| u_i\left(v_i, b_i, b_{\text{-}i} \right) - u_i^{\text{SM}(\lambda)}\left(v_i, b_i, b_{\text{-}i} \right) \right| \mathrm{d} b_{\text{-}i}.
\end{align}
Note that $\det \text{D}\beta_{\text{-} i}^{-1}(b_{\text{-}i})$ is a diagonal matrix, as $\beta_j$ only depends on $v_j$ for every $j\in \mathcal{I}$, so that the determinate is given by the product of the individual inverse functions' derivatives. We continue with bounding the remaining integral. For this, define the set
\begin{align}
	A_{b_i}= \left\{v_{\text{-}i} \in \mathcal{V}_{\text{-}i} \,\Big|\, \max_{\text{-}i} \beta_{\text{-} i} (v_{\text{-}i}) \geq b_i  \right\},
\end{align}
which includes all valuations of bidder $i$'s opponents such that the item is \emph{not} allocated to bidder $i$ in the original game. 
The integral can then be split up in the following way
\begin{align}
	& \int_{\beta_{\text{-} i}\left(\mathcal{V}_{\text{-}i}\right)} \left| u_i\left(v_i, b_i, b_{\text{-}i} \right) - u_i^{\text{SM}(\lambda)}\left(v_i, b_i, b_{\text{-}i} \right) \right| \mathrm{d} b_{\text{-}i}  \\
	&=  \int_{\beta_{\text{-} i}\left(\mathcal{V}_{\text{-}i}\setminus A_{b_i}\right) } \left|  u_i\left(v_i, b_i, b_{\text{-}i} \right) - u_i^{\text{SM}(\lambda)}\left(v_i, b_i, b_{\text{-}i} \right)  \right| \mathrm{d} b_{\text{-}i} && \left(=: (\text{\#}_1) \right)\\
	& + \int_{\beta_{\text{-} i}\left(A_{b_i}\right)} \left|  u_i\left(v_i, b_i, b_{\text{-}i} \right) - u_i^{\text{SM}(\lambda)}\left(v_i, b_i, b_{\text{-}i} \right)  \right| \mathrm{d} b_{\text{-}i} && \left(=: (\text{\#}_2)\right).
\end{align}
It remains to bound the integrals $(\text{\#}_1)$ and $(\text{\#}_2)$. We proceed with $(\text{\#}_2)$, i.e., the integral over the set, where the item is not allocated to bidder $i$. We get
\begin{align}
	(\text{\#}_2) &= \int_{\beta_{\text{-} i}\left(A_{b_i}\right)} \left| u_i^{\text{SM}(\lambda)}\left(v_i, b_i, b_{\text{-}i} \right)  \right| \mathrm{d} b_{\text{-}i}
	= \int_{\beta_{\text{-} i}\left(A_{b_i}\right)} \left| \left( v_i - p^{\text{SM}}(b_i, b_{\text{-}i}) \right) x_i^{\text{SM}(\lambda)}(b_i, b_{\text{-}i}) \right| \mathrm{d} b_{\text{-}i} \\
	\overset{(\text{H}), (\text{A}2.\ref{ass:regularities-interim-bound-pricing}), (\text{A}1.\ref{ass:regularities-interim-bound-val-action-space})}&{\leq} \norm{v_i - p^{\text{SM}}(b_i, \argdot)}_{\infty_{| \beta_{\text{-} i}\left(A_{b_i}\right)}} \cdot \int_{\beta_{\text{-} i}\left(A_{b_i}\right)} \left| x_i^{\text{SM}(\lambda)}(b_i, b_{\text{-}i}) \right| \mathrm{d} b_{\text{-}i}.
\end{align}
The additional subscript of the supremum norm indicates that the domain is limited to $\beta_{\text{-} i}(A_{b_i})$.
This step reduced the problem for $(\text{\#}_2)$ to finding a bound for the integral over the soft-allocation function. Note that the softmax function is strictly positive and strictly decreasing in all components of $b_{\text{-}i}$. If $A_{b_i} = \emptyset$, the integral is zero and any positive number is an upper bound. Otherwise, there exists a $j \neq i$ such that $\beta_j(v_j) \geq b_i$ for all $(v_1, \dots, v_{i-1}, v_{i+1}, \dots, v_j, \dots v_n) \in A_{b_i}$. Therefore, we can bound the integral by
\begin{align}
	\int_{\beta_{\text{-} i}\left(A_{b_i}\right)} \left| x_i^{\text{SM}(\lambda)}(b_i, b_{\text{-}i}) \right| \mathrm{d} b_{\text{-}i} 
	&= \int_{\beta_{\text{-} i}\left(A_{b_i}\right)} \frac{1}{1 + \sum_{j \neq i} \exp\left( \frac{b_j - b_i}{\lambda} \right)} \mathrm{d} b_{\text{-}i} \\
	&\leq \int_{\{b_j \geq b_i\}} \frac{1}{1 + \exp\left(\frac{b_j - b_i}{\lambda} \right)} \mathrm{d} b_j \\
	&\leq \lim_{M \rightarrow  \infty} \int_{b_j = b_i}^{M} \frac{1}{1 + \exp\left(\frac{b_j - b_i}{\lambda} \right)} \mathrm{d} b_j \\
	& = \lim_{M \rightarrow  \infty} \left[ b_j - \lambda \ln \left(\exp\left(\frac{b_j}{\lambda} \right) + \exp\left(\frac{b_i}{\lambda} \right)\right) \right]_{b_j = b_i}^{M} \\
	& = \lim_{M \rightarrow  \infty} M - \lambda \ln\left(\exp\left(\frac{M}{\lambda} \right) + \exp\left(\frac{b_i}{\lambda} \right) \right) - b_i + \lambda \ln \left( 2 \exp\left(\frac{b_i}{\lambda} \right)\right) \\
	&\leq \lim_{M \rightarrow  \infty} M - \lambda \ln\left(\exp\left(\frac{M}{\lambda} \right) \right) - b_i + \lambda \ln \left( 2\right) + \lambda \ln \left( \exp\left(\frac{b_i}{\lambda} \right)\right)  \\
	& = \lambda \ln(2),
\end{align}
which finishes the bound for part $ (\text{\#}_2)$.

We perform similar steps for the integral $ (\text{\#}_1)$, that is taken over the opponents' valuations where bidder $i$ gets the item. Using the definition of the smooth pricing function $p^{\text{SM}}$ over this set, we get
\begin{align}
	(\text{\#}_1) &= \int_{\beta_{\text{-} i}\left(\mathcal{V}_{\text{-}i}\setminus A_{b_i}\right) } \left| v_i - p_i(b_i, b_{\text{-}i}) - \left( v_i - p^{\text{SM}}(b_i, b_{\text{-}i}) \right) x_i^{\text{SM}(\lambda)}(b_i, b_{\text{-}i}) \right| \mathrm{d} b_{\text{-}i} \\
	&= \int_{\beta_{\text{-} i}\left(\mathcal{V}_{\text{-}i}\setminus A_{b_i}\right) } \left| \left( v_i - p_i(b_i, b_{\text{-}i}) \right) \left( 1 - x_i^{\text{SM}(\lambda)}(b_i, b_{\text{-}i}) \right) \right| \mathrm{d} b_{\text{-}i} \\
	\overset{(\text{H}), (\text{A}2.\ref{ass:regularities-interim-bound-pricing})}&{\leq} \norm{v_i - p_i(b_i, \argdot)}_{\infty_{| \beta_{\text{-} i}\left(\mathcal{V}_{\text{-}i} \setminus A_{b_i}\right)}} \cdot \int_{\beta_{\text{-} i}\left(\mathcal{V}_{\text{-}i}\setminus A_{b_i}\right) } \left| 1 - x_i^{\text{SM}(\lambda)}(b_i, b_{\text{-}i}) \right| \mathrm{d} b_{\text{-}i}.
\end{align}

It remains to bound the integral of $h(b_{\text{-}i}) := 1 - x_i^{\text{SM}(\lambda)}(b_i, b_{\text{-}i})$ over the set where bidder $i$ wins the item. Note that $h$ is strictly positive and strictly increasing in all variables $b_{\text{-}i}$. If the set $\mathcal{V}_{\text{-}i} \setminus A_{b_i}$ is empty, we are done. Otherwise, we can bound the integral in the following way:
\begin{align}
	\int_{\beta_{\text{-} i}\left(\mathcal{V}_{\text{-}i}\setminus A_{b_i}\right) } h(b_{\text{-}i}) \, \mathrm{d} b_{\text{-}i}
	&= \int_{\beta_{\text{-} i}\left(\mathcal{V}_{\text{-}i}\setminus A_{b_i}\right) } 1 - \frac{1}{1 + \sum_{j \neq i} \exp \left(\frac{b_j - b_i}{\lambda} \right)} \, \mathrm{d} b_{\text{-}i} \\
	& = \int_{\beta_{\text{-} i}\left(\mathcal{V}_{\text{-}i}\setminus A_{b_i}\right) } \frac{\sum_{j \neq i} \exp \left(\frac{b_j - b_i}{\lambda} \right)}{1 + \sum_{j \neq i} \exp \left(\frac{b_j - b_i}{\lambda} \right)} \, \mathrm{d} b_{\text{-}i} \\
	& \leq \lim_{M \rightarrow - \infty} \int_{M}^{b_i} \cdots \int_{M}^{b_i} \frac{\sum_{j \neq i} \exp \left(\frac{b_j - b_i}{\lambda} \right)}{1 + \sum_{j \neq i} \exp \left(\frac{b_j - b_i}{\lambda} \right)} \mathrm{d} b_1 \dots \mathrm{d} b_{i-1} \mathrm{d} b_{i+1} \dots \mathrm{d} b_n \\ 
	& \leq \lim_{M \rightarrow - \infty} \int_{M}^{b_i} \cdots \int_{M}^{b_i} \sum_{j \neq i} \exp \left(\frac{b_j - b_i}{\lambda} \right) \mathrm{d} b_1 \dots \mathrm{d} b_{i-1} \mathrm{d} b_{i+1} \dots \mathrm{d} b_n \\ 
	& = \lim_{M \rightarrow - \infty} \sum_{j \neq i} \int_{M}^{b_i} \exp \left(\frac{b_j - b_i}{\lambda} \right) \mathrm{d} b_j \\
	&= \lim_{M \rightarrow - \infty} \sum_{j \neq i} \lambda \left[\exp \left(\frac{b_j - b_i}{\lambda} \right) \right]_{b_j = M}^{b_i} \\
	&= \left(n - 1\right) \lambda.
\end{align}
Combining the derived statements, the interim utility error is bounded by
\begin{align}
	\overline{\varepsilon}_i(v_i, b_i, \beta_{\text{-} i}, \lambda) \; \leq \; K(v_i, b_i, \beta_{\text{-} i}) \cdot \lambda,
\end{align}
where
\begin{align}
	K(v_i, b_i, \beta_{\text{-} i}) = 
	&\left(\prod_{j \neq i} \left(\norm{f_j}_{\infty} \cdot \norm{\left(\beta^{-1}_j\right)^{\prime}}_{\infty}\right) \right) \nonumber\\
	&\cdot 
	\left(\ln(2) \, \norm{v_i - p^{\text{SM}}(b_i, \argdot)}_{\infty_{| \beta_{\text{-} i}\left(A_{b_i}\right)}}
	+ \norm{v_i - p(b_i, \argdot)}_{\infty_{| \beta_{\text{-} i}\left(\mathcal{V}_{\text{-}i} \setminus A_{b_i}\right)}} \left(n - 1\right) \right).
\end{align}
Consequently, we can bound the ex ante utility error by performing similar steps as above by
\begin{align}
	\tilde{\varepsilon}_i\left(\beta_i, \beta_{\text{-} i} \right) &:= \left| \int_{\mathcal{V}_{i}} \overline{u}_i\left(v_i, \beta_i(v_i), \beta_{\text{-} i}\right) - \overline{u}_i^{\text{SM}(\lambda)}\left(v_i, \beta_i(v_i), \beta_{\text{-} i} \right) \mathrm{d} F_{i}(v_{i}) \right| \\
	&\leq \lambda \int_{\mathcal{V}_i} K(v_i, \beta_i(v_i), \beta_{\text{-} i}) \, \mathrm{d} F_{i}(v_{i}) \\
	\overset{(\text{H}), (\text{A}2.\ref{ass:regularities-interim-bound-prior}), (\text{A}2.\ref{ass:regularities-interim-bound-strategies}), (\text{A}2.\ref{ass:regularities-interim-bound-inverse})}&{\leq} \tilde{K} \cdot \lambda,
\end{align}
where 
\begin{align}
	\tilde{K} = \mu\left(\beta_i\left(\mathcal{V}_i \right) \right) \left(\prod_{j \in \mathcal{I}} \norm{f_j}_{\infty} \cdot \norm{\left(\beta^{-1}_j\right)^{\prime}}_{\infty} \right)  \left( \ln(2) \, \norm{g_i^{\text{SM}}}_{\infty} + \left(n-1 \right) \norm{g_i}_{\infty} \right),
\end{align}
where $\mu$ denotes the Borel-measure and for the functions
\begin{align}
	g_i^{\text{SM}}(b_i, \beta_i^{-1}(b_i)) &= \norm{\beta_i^{-1}(b_i) - p^{\text{SM}}(b_i, \argdot)}_{\infty_{| \beta_{\text{-} i}\left(A_{b_i}\right)}},\\
	g_i(b_i, \beta_i^{-1}(b_i)) &= \norm{\beta_i^{-1}(b_i) - p_i(b_i, \argdot)}_{\infty_{| \beta_{\text{-} i}\left(\mathcal{V}_{\text{-}i} \setminus A_{b_i}\right)}}.
\end{align}

\end{proof}

\begin{remark}
	The proof uses the infinity norm of the H\"older inequality $\left(\norm{fg}_1 \leq \norm{f}_1 \norm{g}_{\infty} \right)$, so that one needs Condition~\ref{ass:regularities-interim-bound-inverse} of Assumption~\ref{ass:regularities-for-interim-bound}. One can use the $\norm{ \cdot}_2$-norm in the inequality $\left(\norm{fg}_1 \leq \norm{f}_2 \norm{g}_2 \right)$, so that one can weaken the assumption to $\norm{\left(\beta_{i}^{-1}\right)^{\prime}}_2 < \infty$. However, this leads to a worse convergence rate of the form $\mathcal{O}(\sqrt{\lambda})$.
\end{remark}

\begin{remark}
	The result also holds for a special case of risk-averse bidders. Assume the ex post utility functions additionally depend on a risk parameter $\rho \in (0, 1]$:
	\begin{align}
		u_i(v_i, b, \rho) \; &= \; \left(v_i \, x_i(b) - p_i(b)\right)^{\rho} \\
		u_i^{\text{SM}(\lambda)}(v_i, b, \rho) &= \left(\left(v_i - p^{\text{SM}}(b)\right) x_i^{\text{SM}(\lambda)}(b)\right)^{\rho}.
	\end{align}
	By using Jensen's inequality for the concave mapping $x\mapsto x^{\rho}$, one can derive a convergence rate of $\mathcal{O}(\lambda^\rho)$ by performing analogous steps as in the proof above.
\end{remark}

\section{Proof of Theorem~\ref{thm:eps-equ-in-smooth-game-to-original-game}}\label{sec:proof-cor-1}
Let us now prove that an $\varepsilon$-equilibrium in the smoothed game translates to an $\varepsilon + \mathcal{O}(\lambda)$-equilibrium in the original auction.

\begin{proof}
	We know there exists a constant $K$ such that $\abs{\tilde{u}_i^{\text{SM}(\lambda)} (\beta) - \tilde{u}_i(\beta)} \leq K \lambda$ due to Proposition~\ref{thm:linear-convergence-utility-error}. Therefore, the following holds for an $\varepsilon$-BNE in the smoothed game $\beta^*$ and any strategy $\beta_i$:
	\begin{align}
		\tilde{u}_i(\beta_i, \beta_{-i}^*) - \tilde{u}_i (\beta_i^*, \beta_{-i}^*) & = \tilde{u}_i(\beta_i, \beta_{-i}^*) - \tilde{u}_i^{\text{SM}(\lambda)} (\beta_i, \beta_{-i}^*) + \tilde{u}_i^{\text{SM}(\lambda)} (\beta_i, \beta_{-i}^*) \nonumber\\
		& - \tilde{u}_i^{\text{SM}(\lambda)} (\beta_i^*, \beta_{-i}^*) + \tilde{u}_i^{\text{SM}(\lambda)} (\beta_i^*, \beta_{-i}^*) - \tilde{u}_i (\beta_i^*, \beta_{-i}^*) \\
		& \leq K \lambda + \varepsilon + K \lambda = \varepsilon + 2K\lambda.
	\end{align}
	This is equivalent to
	\begin{align}
		\tilde \ell_i(\beta_i^*, \beta_{-i}^*) \leq \varepsilon + 2K \lambda.
	\end{align}
\end{proof}

\section{A Special Case: Exact Errors}
\label{sec:exact-interim-error}

Consider the single-item FPSB auction with two bidders having uniform priors on the unit interval. Suppose that bidder~$2$ has a linear strategy $\beta_2(v_2) = s \, v_2$ with $s \in (0, 1]$. Then we can derive an error rate for bidder~$1$'s absolute interim utility difference $\overline{\varepsilon}_\lambda = |\overline{u}_1 - \overline{u}_1^\text{SM}|$ and for valuation $v_1$ and bid $b_1 \leq s$.

For the interim utility of bidder~$1$ in the original game, we have
\begin{align}
	\overline{u}_1(v_1, b_1) = \int_{v_2=0}^\frac{b_1}{s} (v_1-b_1) \, \mathrm{d}v_2 = \frac{v_1 - b_1}{s} \, b_1.
\end{align} 
In the smoothed auction, we have
\begin{align}
	\overline{u}_1^\text{SM}(v_1, b_1) = \int_{v_2=0}^1 (v_1 - \max\{b_1, s \, v_2\}) x_1^\text{SM} (b_1, s \, v_2) \, \mathrm{d}v_2.
\end{align} 
When splitting the domain of the integral at $b_1 = s \, v_2$, the first integral evaluates to
\begin{align*}
	-\dfrac{b_1 - v_1}{s} \left({\lambda}\ln\left(\mathrm{e}^\frac{b}{{\lambda}}+1\right)-{\lambda}\ln\left(2\mathrm{e}^\frac{b_1}{{\lambda}}\right)+b_1\right)
\end{align*}
and the second to
\begin{align*}
	&\dfrac{1}{2s} \Big[ 2s{\lambda}v_2\ln\left(\mathrm{e}^{\frac{sv_2}{{\lambda}}-\frac{b_1}{{\lambda}}}+1\right)+2{\lambda}^2\operatorname{Li}_2\left(-\mathrm{e}^{\frac{sv_2}{{\lambda}}-\frac{b_1}{{\lambda}}}\right)-2v_1{\lambda}\ln\left(\mathrm{e}^\frac{sv_2}{{\lambda}}+\mathrm{e}^\frac{b_1}{{\lambda}}\right)-s^2v_2^2+2v_1sv_2 \Big]_{v_2 = \frac{b_1}{s}}^1\\
	&\quad= \dfrac{1}{s} \Big(\lambda (s\ln(\mathrm{e}^{s/\lambda - b_1/\lambda} + 1) - b_1\ln(2)) + \lambda^2 (\operatorname{Li}_2(-\mathrm{e}^{(s-b_1)/\lambda}) + \frac{1}{12}\pi^2)\\
	&\quad + v_1\lambda (\ln(2) + b_1/\lambda - \ln(\mathrm{e}^{s/\lambda} + \mathrm{e}^{b_1/\lambda})) + \frac{1}{2}(b_1^2 - s^2) + v_1(s-b_1) \Big).
\end{align*}
Combining these results, we arrive at an exact interim error of
\begin{align}
	\label{eq:interim-error}
	\overline{\varepsilon}_\lambda(v_1, b_1) &= \frac{\lambda}{s} \Bigg( - s \ln\left(\mathrm{e}^{\frac{s-b_1}{\lambda}} + 1\right) - \lambda \left(\operatorname{Li}_2\left(-\mathrm{e}^{\frac{s-b_1}{\lambda}}\right) + \frac{1}{12}\pi^2\right)\nonumber\\
	&\quad + v_1 \ln(\mathrm{e}^{s/\lambda} + \mathrm{e}^{b_1/\lambda}) + \frac{1}{\lambda} \left(\frac{s^2 - b_1^2}{2} - v_1s\right)\nonumber\\
	&\quad + (b_1-v_1) \left(\ln(\mathrm{e}^{-b_1/\lambda} + 1)\right)\Bigg).
\end{align}
Here, $\operatorname{Li}_2$ is the dilogarithm.
This result shows vastly different convergence rates across the valuation and action space.
Let us assume both bidders are playing according to their BNE strategy, $\beta_i(v_i) = 0.5 v_i$. Now, for the extreme case of $v_1 = 1$, $\overline{\varepsilon}_\lambda$ tends towards a linear function. At the other end of the spectrum, for $v_1 = 0.5$, $\overline{\varepsilon}_\lambda$ tends towards being constantly zero. In summary, the higher the valuation is the slower the convergence rate, with a linear rate in the worst case.
Figure~\ref{fig:smooth_error_bound} in the main paper shows the convergence rates for smaller temperatures. We depict a selection of three valuations. 

Again assuming that both bidders are playing according to their BNE strategy, the ex ante error
\begin{equation}
	\tilde{\varepsilon}_\lambda(\beta_1) \; \vcentcolon= \; \int_{v_1=0}^1 \overline{\varepsilon}_\lambda(v_1, \beta_1(v_1)) \, \mathrm{d}v_1
\end{equation}
can be approximated by taking the sample mean of Equation~\ref{eq:interim-error}. The expression goes towards zero quickly for ever smaller temperatures and is depicted in Figure~\ref{fig:smooth_error_bound} in the main paper.

\section{Impact of Batch Size}\label{sec:bacth-size}
We have run experiments for different batch sizes. The performance increase for ever larger batch sizes diminishes and optimal results are reached for temperature values just below $0.01$ as can be seen in Table~\ref{tab:batch_size}.

\begin{table}[t]
	\centering
	\begin{tabular}{lrr}
		\toprule
		Batch size               & $\min L_2$ & $\lambda$ \\
		\midrule
		$2^{10} = 1{,}024$       & 0.0177     & 0.0239    \\
		$2^{14} = 16{,}384$      & 0.0044     & 0.0119    \\
		$2^{18} = 262{,}144$     & 0.0044     & 0.0089    \\
		$2^{22} = 4{,}194{,}304$ & 0.0042     & 0.0089    \\
		\bottomrule
	\end{tabular}
	\caption{Results for a selection of different batch sizes in the FPSB base setting. We report $\lambda$ for which the $L_2$ loss is minimized. $2^{18}$ is the default value used during training and $2^{22}$ the maximal value possible on the GPU used.}
	\label{tab:batch_size}
\end{table}

\section{Reproducibility and Hyperparameters\label{sec:reproducibility}}
We have implemented all auctions and algorithms in the PyTorch framework. The code is available at \href{https://github.com/heidekrueger/bnelearn}{Github}.

\subsection{Learning}
We use common hyperparameters across all settings except where noted otherwise. 
The feed-forward neural networks are fully connected with two hidden layers of ten nodes each with SeLU activations, as well as ReLU activations applied to the output layer. 
We model all bidders by a shared policy because the auctions considered are symmetric. Hence, learning is stabilized but limited to finding symmetric BNE. Furthermore, we perform supervised pretraining of 50 iterations towards truthful strategies to prevent degenerate initializations.
All experiments are run on a single Nvidia GeForce 2080Ti GPU with 11\,GB of memory and a batch size of $2^{18}$ for learning.
Each experiment was repeated five times with 2{,}000 iterations.
Furthermore, the following algorithm specific settings were used:
\begin{itemize}
	\item  For NPGA, we choose a population size of 64 and a variance of 1 for the normal distribution from which we draw population samples in parameter space. The variance is then scaled by the model size as is done in \citep{bichler2021learning}.
	\item In the case of REINFORCE, the output dimension is increased by a factor of two because for each bid, a normal distribution (with its two parameters) is learned instead.
	\item For the smoothed game, we choose a temperature of 0{.}01.
\end{itemize}

\subsection{Evaluating}
A batch size of $2^{22}$ was used for the calculation of the $L_2$ loss. The choice of batch sizes was mainly driven by maxing out the GPU memory. Learning requires more memory than evaluating $L_2$, so the latter was possible to conduct with larger batch sizes.
For the utility loss $\varepsilon$, we decreased the number of prior samples from the player currently under evaluation to $n_\text{own-batch} = 2^{10}$. For each of these valuations, his or her best response --- with possible actions from an equidistant grid of size $n_\text{gird} = 2^{10}$ --- is approximated over a sample of $n_\text{opponent-batch} = 2^{20}$ opponent valuations. A higher batch size for the opponents is necessary for reaching the required precision in estimating the utilities. In total, the calculation of $\varepsilon$ requires $n_\text{own-batch} \cdot n_\text{gird} \cdot n_\text{opponent-batch} = 2^{40} > 1\,$trillion game evaluations.

%% file: ms.bbl
\begin{thebibliography}{23}
\providecommand{\natexlab}[1]{#1}
\providecommand{\url}[1]{\texttt{#1}}
\expandafter\ifx\csname urlstyle\endcsname\relax
  \providecommand{\doi}[1]{doi: #1}\else
  \providecommand{\doi}{doi: \begingroup \urlstyle{rm}\Url}\fi

\bibitem[Armantier et~al.(2008)Armantier, Florens, and
  Richard]{armantier2008approximation}
Armantier, O., Florens, J.-P., and Richard, J.-F.
\newblock Approximation of {N}ash equilibria in {B}ayesian games.
\newblock \emph{Journal of Applied Econometrics}, 23\penalty0 (7):\penalty0
  965--981, 2008.

\bibitem[Athey(2001)]{athey2001single}
Athey, S.
\newblock Single crossing properties and the existence of pure strategy
  equilibria in games of incomplete information.
\newblock \emph{Econometrica}, 69\penalty0 (4):\penalty0 861--889, 2001.

\bibitem[Bangaru et~al.(2021)Bangaru, Michel, Mu, Bernstein, Li, and
  Ragan-Kelley]{bangaru2021systematically}
Bangaru, S.~P., Michel, J., Mu, K., Bernstein, G., Li, T.-M., and Ragan-Kelley,
  J.
\newblock Systematically differentiating parametric discontinuities.
\newblock \emph{ACM Transactions on Graphics (TOG)}, 40\penalty0 (4):\penalty0
  1--18, 2021.

\bibitem[Bhawalkar \& Roughgarden(2011)Bhawalkar and
  Roughgarden]{bhawalkar2011welfare}
Bhawalkar, K. and Roughgarden, T.
\newblock Welfare guarantees for combinatorial auctions with item bidding.
\newblock In \emph{Proceedings of the twenty-second annual ACM-SIAM symposium
  on Discrete Algorithms}, pp.\  700--709. SIAM, 2011.

\bibitem[Bichler et~al.(2021)Bichler, Fichtl, Heidekr{\"u}ger, Kohring, and
  Sutterer]{bichler2021learning}
Bichler, M., Fichtl, M., Heidekr{\"u}ger, S., Kohring, N., and Sutterer, P.
\newblock Learning equilibria in symmetric auction games using artificial
  neural networks.
\newblock \emph{Nature Machine Intelligence}, 3\penalty0 (8):\penalty0
  687--695, 2021.

\bibitem[Bogachev(2007)]{bogachev2007measure}
Bogachev, V.~I.
\newblock \emph{Measure Theory}, volume~1.
\newblock Springer Science \& Business Media, 2007.
\newblock URL \url{https://doi.org/10.1007/978-3-540-34514-5}.

\bibitem[Bosshard et~al.(2020)Bosshard, B{\"u}nz, Lubin, and
  Seuken]{bosshardComputingBayesNashEquilibria2020}
Bosshard, V., B{\"u}nz, B., Lubin, B., and Seuken, S.
\newblock Computing bayes-nash equilibria in combinatorial auctions with
  verification.
\newblock \emph{Journal of Artificial Intelligence Research}, 69:\penalty0
  531--570, 2020.

\bibitem[Chasnov et~al.(2020)Chasnov, Ratliff, Mazumdar, and
  Burden]{chasnov2020convergence}
Chasnov, B., Ratliff, L., Mazumdar, E., and Burden, S.
\newblock Convergence analysis of gradient-based learning in continuous games.
\newblock In \emph{Uncertainty in Artificial Intelligence}, pp.\  935--944.
  PMLR, 2020.

\bibitem[Huang et~al.(2021)Huang, Hu, Du, Zhou, Su, Tenenbaum, and
  Gan]{huang2021plasticinelab}
Huang, Z., Hu, Y., Du, T., Zhou, S., Su, H., Tenenbaum, J.~B., and Gan, C.
\newblock Plasticinelab: A soft-body manipulation benchmark with differentiable
  physics.
\newblock In \emph{International Conference on Learning Representations}, 2021.
\newblock URL \url{https://openreview.net/forum?id=xCcdBRQEDW}.

\bibitem[Katz(1982)]{katzChangeVariablesMultiple1982}
Katz, V.~J.
\newblock Change of variables in multiple integrals: Euler to cartan.
\newblock \emph{Mathematics Magazine}, 55:\penalty0 3--11, 1982.
\newblock ISSN 0025-570X.
\newblock URL \url{https://www.jstor.org/stable/2689856}.

\bibitem[Krishna(2009)]{krishna2009auction}
Krishna, V.
\newblock \emph{Auction theory}.
\newblock Academic press, 2009.
\newblock URL \url{https://doi.org/10.1016/C2009-0-22474-3}.

\bibitem[Letcher(2020)]{letcher2020impossibility}
Letcher, A.
\newblock On the impossibility of global convergence in multi-loss
  optimization.
\newblock In \emph{International Conference on Learning Representations}, 2020.

\bibitem[Li \& Wellman(2021)Li and Wellman]{li2021evolution}
Li, Z. and Wellman, M.~P.
\newblock Evolution strategies for approximate solution of bayesian games.
\newblock In \emph{Proceedings of the AAAI Conference on Artificial
  Intelligence}, volume~35, pp.\  5531--5540, 2021.

\bibitem[Mazumdar et~al.(2020)Mazumdar, Ratliff, Jordan, and
  Sastry]{mazumdar2020policy}
Mazumdar, E., Ratliff, L.~J., Jordan, M.~I., and Sastry, S.~S.
\newblock Policy-gradient algorithms have no guarantees of convergence in
  linear quadratic games.
\newblock In \emph{Proceedings of the 19th International Conference on
  Autonomous Agents and MultiAgent Systems}, pp.\  860--868, 2020.

\bibitem[Mohamed et~al.(2020)Mohamed, Rosca, Figurnov, and
  Mnih]{mohamed2020monte}
Mohamed, S., Rosca, M., Figurnov, M., and Mnih, A.
\newblock Monte carlo gradient estimation in machine learning.
\newblock \emph{Journal of Machine Learning Research}, 21\penalty0
  (132):\penalty0 1--62, 2020.

\bibitem[Ott \& Beck(2013)Ott and Beck]{ott2013incentives}
Ott, M. and Beck, M.
\newblock Incentives for overbidding in minimum-revenue core-selecting
  auctions.
\newblock 2013.

\bibitem[Salimans et~al.(2017)Salimans, Ho, Chen, Sidor, and
  Sutskever]{salimans2017evolution}
Salimans, T., Ho, J., Chen, X., Sidor, S., and Sutskever, I.
\newblock Evolution {{Strategies}} as a {{Scalable Alternative}} to
  {{Reinforcement Learning}}.
\newblock \emph{ArXiv}, March 2017.
\newblock URL \url{https://arxiv.org/abs/1703.03864}.

\bibitem[Schulman et~al.(2017)Schulman, Wolski, Dhariwal, Radford, and
  Klimov]{schulman2017proximal}
Schulman, J., Wolski, F., Dhariwal, P., Radford, A., and Klimov, O.
\newblock Proximal policy optimization algorithms.
\newblock \emph{ArXiv}, abs/1707.06347, 2017.

\bibitem[Suh et~al.(2022)Suh, Simchowitz, Zhang, and
  Tedrake]{suh2022differentiable}
Suh, H.~J., Simchowitz, M., Zhang, K., and Tedrake, R.
\newblock Do differentiable simulators give better policy gradients?
\newblock In \emph{International Conference on Machine Learning}, pp.\
  20668--20696. PMLR, 2022.

\bibitem[Sutton \& Barto(2018)Sutton and Barto]{sutton2018reinforcement}
Sutton, R.~S. and Barto, A.~G.
\newblock \emph{Reinforcement learning: An introduction}.
\newblock MIT press, 2018.

\bibitem[Talvila(2001)]{talvilaNecessarySufficientConditions2001}
Talvila, E.
\newblock Necessary and sufficient conditions for differentiating under the
  integral sign.
\newblock \emph{The American Mathematical Monthly}, 108\penalty0 (6):\penalty0
  544--548, 2001.
\newblock URL \url{https://www.jstor.org/stable/2695709}.

\bibitem[Vickrey(1961)]{vickrey1961counterspeculation}
Vickrey, W.
\newblock Counterspeculation, auctions, and competitive sealed tenders.
\newblock \emph{The Journal of finance}, 16\penalty0 (1):\penalty0 8--37, 1961.

\bibitem[Waugh et~al.(2009)Waugh, Schnizlein, Bowling, and
  Szafron]{waugh2009abstraction}
Waugh, K., Schnizlein, D., Bowling, M.~H., and Szafron, D.
\newblock Abstraction pathologies in extensive games.
\newblock \emph{AAMAS (2)}, 2009:\penalty0 781--8, 2009.

\end{thebibliography}
